\documentstyle[12pt,html,psfig]{article}

\begin{document}
\title{MATTERS OF GRAVITY, The newsletter of the APS Topical Group on 
Gravitation}
\begin{center}
{ \Large {\bf MATTERS OF GRAVITY}}\\ 
\bigskip
\hrule
\medskip
{The newsletter of the Topical Group on Gravitation of the American Physical 
Society}\\
\medskip
{\bf Number 20 \hfill Fall 2002}
\end{center}
\begin{flushleft}

\tableofcontents
\vfill
\section*{\noindent  Editor\hfill}

Jorge Pullin\\
\smallskip
Department of Physics and Astronomy\\
Louisiana State University\\
202 Nicholson Hall\\
Baton Rouge, LA 70803-4001\\
Phone/Fax: (225)578-0464\\
Internet: 
\htmladdnormallink{\protect {\tt{pullin@phys.lsu.edu}}}
{mailto:pullin@phys.lsu.edu}\\
WWW: \htmladdnormallink{\protect {\tt{http://www.phys.lsu.edu/faculty/pullin}}}
{http://www.phys.lsu.edu/faculty/pullin}\\
\hfill ISSN: 1527-3431
\begin{rawhtml}
<P>
<BR><HR><P>
\end{rawhtml}
\end{flushleft}
\pagebreak
\section*{Editorial}

Not much to report here. This newsletter ended a bit long, due to
the large number of conference reports, just reflecting the increasing
number of meetings that are taking place in our field. It is also
a bit late, also due to the large number of articles involved. I
apologize for this.

The next newsletter is due February 1st.
If everything goes well this newsletter should be available in the
gr-qc arXiv archives 
(\htmladdnormallink{{\tt http://www.arxiv.org}}{http://www.arxiv.org})
under number gr-qc/0209085. To retrieve it
send email to
\htmladdnormallink{gr-qc@arxiv.org}{mailto:gr-qc@arxiv.org}
with Subject: get 0209085
(numbers 2-19 are also available in gr-qc). All issues are available in the
WWW:\\\htmladdnormallink{\protect {\tt{http://www.phys.lsu.edu/mog}}}
{http://www.phys.lsu.edu/mog}\\ 

The newsletter is  available for
Palm Pilots, Palm PC's and web-enabled cell phones as an
Avantgo channel. Check out 
\htmladdnormallink{\protect {\tt{http://www.avantgo.com}}}
{http://www.avantgo.com} under technology$\rightarrow$science.

A hardcopy of the newsletter is
distributed free of charge to the members of the APS
Topical Group on Gravitation upon request (the default distribution form is
via the web) to the secretary of the Topical Group. 
It is considered a lack of etiquette to
ask me to mail you hard copies of the newsletter unless you have
exhausted all your resources to get your copy otherwise.
\par
If you have comments/questions/complaints about the newsletter email
me. Have fun.
\bigbreak

\hfill Jorge Pullin\vspace{-0.8cm}
\section*{Correspondents}
\begin{itemize}
\item John Friedman and Kip Thorne: Relativistic Astrophysics,
\item Raymond Laflamme: Quantum Cosmology and Related Topics
\item Gary Horowitz: Interface with Mathematical High Energy Physics and
String Theory
\item Beverly Berger: News from NSF
\item Richard Matzner: Numerical Relativity
\item Abhay Ashtekar and Ted Newman: Mathematical Relativity
\item Bernie Schutz: News From Europe
\item Lee Smolin: Quantum Gravity
\item Cliff Will: Confrontation of Theory with Experiment
\item Peter Bender: Space Experiments
\item Riley Newman: Laboratory Experiments
\item Warren Johnson: Resonant Mass Gravitational Wave Detectors
\item Gary Sanders: LIGO Project
\item Peter Saulson: former editor, correspondent at large.
\end{itemize}
\vfill
\pagebreak

\section*{\centerline {
Einstein Prize update}}
\addtocontents{toc}{\protect\medskip}
\addtocontents{toc}{\bf TGG News:}
\addcontentsline{toc}{subsubsection}{\it  
Einstein prize update, by Clifford Will}
\begin{center}
Clifford Will, Washington University,  St. Louis
\htmladdnormallink{cmw@wuphys.wustl.edu}
{mailto:cmw@wuphys.wustl.edu}
\end{center}

The American Physical Society has formally established the Einstein Prize
in Gravitational Physics as a biennial prize, for outstanding achievement
in theory, experiment or observation in gravitational physics.
Fundraising for the prize continues, however, with the goal of reaching an
endowment of \$200 K, so that the prize could be awarded annually.  David
Lee continues to match every dollar raised for the prize.  For
information, contact Clifford Will (
\htmladdnormallink{cmw@wuphys.wustl.edu}
{mailto:cmw@wuphys.wustl.edu}
) or check
\htmladdnormallink{http://wugrav.wustl.edu/People/CLIFF/prize.html}
{http://wugrav.wustl.edu/People/CLIFF/prize.html}.  A prize selection
committee was established by the APS, with the intention of awarding the
initial prize in 2003, at the April APS meeting.  To nominate somebody for
the prize, follow the guidelines that can be found at
\htmladdnormallink{http://www.aps.org/praw/nomguide.html}
{http://www.aps.org/praw/nomguide.html}.

\bigskip

\section*{\centerline {
World year of physics 2005}}
\addcontentsline{toc}{subsubsection}{\it  
World year of physics, by Richard H. Price}
\begin{center}
Richard H. Price, University of Utah
\htmladdnormallink{rprice@mail.physics.utah.edu}
{mailto:rprice@mail.physics.utah.edu}
\end{center}

The year 2005 will be the centenary year of the 1905 Einstein miraculous
year, in connection with this, UNESCO has declared 2005 a World Year of
Physics.  Each country would plan its own events and perhaps there will be
an international meeting in Switzerland. The European Physical Society is
leading the way and heading the European side of things.

The APS plans to be part of this, and they are thinking about outreach
activities under the tentative title ``Einstein in the 21st Century."  
Although Einstein's 1905 achievements really had nothing to do with the G
of TGG, we do have a special claim on Einstein (the Einstein prize, for
example). The role we will play in 2005 is now being considered.

\vfill
\pagebreak

\section*{\centerline {
We hear that...}}
\addcontentsline{toc}{subsubsection}{\it  
We hear that... by Jorge Pullin}
\begin{center}
Jorge Pullin
\htmladdnormallink{pullin@phys.lsu.edu}
{mailto:pullin@phys.lsu.edu}
\end{center}

{\bf Robert Wald} has been elected to the National Academy of Sciences
(I incorrectly reported this in the fall 2001 issue, at that time he
had been elected to the American Academy of Arts and Sciences). 

{\bf Matt Choptuik} was chosen among the top 20 researchers 40 years of
age and under by the Canadian Insitute for Advanced Research (CIAR).

{\bf Patrick Brady and Amanda Peet} have been awarded Sloan Fellowships.
Brady also received a a Cottrell Science award from the Research Corporation.

{\bf Raphael Bousso and Fotini Markopoulou} were chosen as the winners of
the young researchers competition associated with the ``Science and
ultimate reality'' symposium in honor of John Archibald Wheeler.

{\bf The Perimeter Institute} in Waterloo Canada has received a major
commitment from the Canadian Government. It will contribute \$25
million over five years. Additional \$11 million will be contributed
towards the building. Prime Minister Chr\'etien spoke at the ground-breaking
ceremony for the new building.

Hearty congratulations!

\vfill
\pagebreak
\section*{\centerline {
R-mode epitaph?}}
\addtocontents{toc}{\protect\medskip}
\addtocontents{toc}{\bf Research briefs:}
\addcontentsline{toc}{subsubsection}{\it  
R-mode epitaph? by John Friedman and Nils Andersson}
\begin{center}
John Friedman, UW-Milwaukee, Nils Andersson, Univ. of Southampton UK
\htmladdnormallink{friedman@thales.phys.uwm.edu}
{mailto:friedman@thales.phys.uwm.edu},
\htmladdnormallink{na@maths.soton.ac.uk}
{mailto:na@maths.soton.ac.uk}

\end{center}

For the past few years, because the growth time of the
gravitational-wave driven instability (CFS instability) is short for
r-modes, it has seemed likely that the instability limits the spin of
rapidly rotating, newly formed neutron stars - and possible that it
limits the spin of old, accreting neutron stars as well.  Because most
of the initial angular momentum of the star could be radiated, the
gravitational waves emitted by sources as close as the Virgo Cluster
might be observable by detectors with the expected sensitivity of
advanced LIGO. In the past year, however, studies of two
mechanisms -- rapid cooling and high viscosity associated with a
hyperons in neutron-star cores, and amplitude saturation by nonlinear
coupling to stable modes have sharply decreased the likelihood that the
waves from nascent neutron stars will be detectable.   Whether the
instability limits the spin of neutron stars remains an open
question.  And, surprisingly, the hyperon studies may have increased
the chance of detecting gravitational waves from old accreting
stars.\\

One of the major  questions about the r-mode instability was whether
nonlinear couplings would sharply limit its amplitude.  The first
studies of this problem were numerical evolutions of the nonlinear
equations:  First, Stergioulas and Font [1] carried out a numerical
evolution of the exact relativistic Euler equations on a background
spacetime, with the initial data of a large-amplitude r-mode.  They
found, surprisingly, that there was no apparent energy transfer to
daughter modes until the perturbation's amplitude was substantially
larger than unity.  Next was a numerical time-evolution of the full
Newtonian perturbation equations by Lindblom, Tohline and Vallisneri
[2]. Because the actual growth time from radiation-reaction is
impractically long, they looked at the maximum amplitude for a
perturbation driven by a greatly enlarged radiation-reaction term. They
again found saturation only at amplitude large compared to unity; the
limit appeared to be set by a shock wave, a dramatic breaking on the
star's surface.  \\

A clear implication of these numerical evolutions was that coupling to
low-order modes did not set a stringent limit on the r-mode amplitude
-- not, at least, within a time of about ten rotational periods, and with
the small initial daughter-mode amplitudes and numerical viscosity of the
simulations.  These evolutions can examine the full nonlinear coupling,
but they were limited in their resolution and evolution time.  The
complementary way to address the question is in 2nd-order perturbation
theory - by omitting higher-order couplings, one reduces the nonlinear
evolution to a set of coupled ordinary differential equations.  But the
second-order perturbation theory of a rotating star had not been
developed, and it is was a major undertaking, completed this year by
Arras, Flanagan, Morsink, Schenk, Teukolsky, and Wasserman [3].  The
implication of their work is opposite to the indications of the fully
nonlinear evolutions:  Nonlinear couplings sharply limit the amplitude
of an unstable r-mode.\\

Their work is consistent with the fully nonlinear numerical evolutions,
because the coupling they find to low-order modes is consistent with a
saturation amplitude of order unity.  It is, instead, the coupling to
many short-wavelength modes with frequencies comparable to that of the
r-mode, that saturates the mode at an amplitude smaller than
$10^{-2}$.   (And the most recent fully nonlinear numerical evolution,
by Gressman et al.[4], finds that increasing the resolution yields
rapid nonlinear decay of the mode at decreased amplitude.)  Nevertheless,
r-modes may set the limit on rotation of young rapidly rotating neutron
stars (this is less likely if the actual saturation amplitude is
$10^{-4}$); and the r-mode instability may account for the maximum
rotation period observed in old neutron stars spun-up by accretion
(i.e., in x-ray binaries). \\

Exotic particles in the core of a neutron star lead to a significantly
stronger viscous damping than assumed in the first studies of the
unstable r-modes.  Of particular relevance is the presence of hyperons
and non-leptonic weak reactions. Hyperon-mediated damping of
neutron-star oscillations was proposed by Langer and Cameron in 1969
and by Peter Jones in 1970, but the minimum density expected for
hyperons was apparently $10^{15}$ g/cm$^3$ (see, for example,
Zel'dovich and Novikov, v. 1, Sect. 11.5, in paragraphs written by
Thorne).  Spurred by Jones' recent reminder [5] that the r-mode
instability  would be completely suppressed in stars with a
significant hyperon fraction, Lindblom and Owen [6], in a
tour-de-force calculation using fully relativistic cross sections,
found the bulk viscosity coefficient for non-leptonic reactions.
Independently Haensel, Levenfish and Yakovlev [7] considered
superfluid hyperons and obtained a set of approximate formulas for
various bulk viscosity coefficients.  These results show that, when
hyperons are present, the dissipation due to direct URCA is
overwhelming. In fact, the hyperon cooling is so rapid that, even
without the enhanced hyperon bulk viscosity, no mode of a nascent
neutron star would have time to grow before it was stabilized by
viscous damping.\\

There is, however, significant uncertainty in the critical density at
which hyperons appear and in the central density of neutron stars.
Even for the equation of state (due to Glendenning) that Lindblom and
Owen consider, the central density of a 1.4 M$_\odot$ star is near
that needed for hyperons to appear
when the rotation  reaches its maximum
value.  The averaged measured mass of neutron stars is somewhat smaller
than this; and the compressibility consistent with our knowledge of
matter above nuclear density ranges over a factor of 5 or more for a
1.4 M$_\odot$ star, an error bar easily large enough to prevent
our knowing whether or not neutron stars have hyperon cores.\\

The possibility that gravitational waves from unstable modes could
balance the accretion torque, and hence halt the associated spin-up,
was first discussed by Papaloizou and Pringle and Wagoner for
the f-modes.  The analogous scenario for the r-modes was analyzed in
detail by Andersson, Kokkotas and Stergioulas and  Bildsten
a few years ago.  Should this happen, neutron stars in LMXBs would be
promising
sources for detectable gravitational waves.  In independent work,
Spruit and Levin pointed  out that this scenario may not work:  The
unstable mode will heat the star up (via the shear viscosity), and, if
the viscosity gets weaker as the temperature increases, the
mode-heating will trigger a rapid thermal runaway during which the star
will spin down.  This mechanism would seem to rule out r-modes in
galactic LMXBs as a source of detectable gravitational waves, because
they will only radiate for a tiny fraction of the system's lifetime.\\

Interestingly, the recent results indicating a small saturation
amplitude and strong ``exotic'' bulk viscosity may imply that r-modes
in these systems would be relevant gravitational-wave sources after
all.  Very recently, Wagoner [8] reported work showing that old,
accreting neutron stars with a hyperon core  will not undergo a thermal
runaway, but will reach a the quasi-equilibrium state in which
gravitational wave emission balances accretion spin-up {\em and cooling
balances shear viscosity heating}.  Andersson, Jones and Kokkotas [9] had
previously found the same balance for strange stars, and it might also
characterize accreting neutron stars with quark cores. Most
importantly, in these scenarios the r-mode amplitude required to
balance the accretion torque is orders of magnitude smaller than the
saturation amplitude estimated by Arras et al.  Finally, Heyl [10] has
investigated the effect of a small saturation amplitude on the thermal
runaway in a ``normal'' neutron star. His results show that the phase
during which the r-modes radiate gravitationally is significantly
extended if the modes cannot grow to a large amplitude (as long as the
modes are not wiped out by the saturation mechanism). \\

One must place hyperon damping and nonlinear saturation at the top of
the list of mechanisms that may kill or maim unstable r-modes [11], but
reports of their demise may be premature. \\

\par
{\em{References:}}
\par
1. N. Stergioulas and J.A. Font, Phys. Rev. Lett {\bf 86} 1148 (2001)
\par
2. L. Lindblom, J.E. Tohline and M. Vallisneri Phys. Rev. Lett {\bf 86} 1152
(2001); Phys. Rev. D {\bf 65} 084039 (2002)

\par
3.  P. Arras, E.E. Flanagan, S.M. Morsink, A.K. Schenk, S.A. Teukolsky
and I. Wasserman, {\em Saturation of the r-mode instability} preprint
\htmladdnormallink{astro-ph/0202345}{http://arxiv.org/abs/astro-ph/0202345}
; S.M. Morsink, Ap.J. {\bf 571}, 435 (2002);
A.K. Schenk, P. Arras, E.E. Flanagan, S.A. Teukolsky and I. Wasserman,
Phys. Rev. D {\bf 65}, 024001 (2002).
\par
4. P. Gressman, L.-M. Lin, W.-M. Suen, N. Stergioulas and J.~L. Friedman,
Phys. Rev. D {\bf 66}, 041303 (2002).
\par
5. P.B. Jones, Phys. Rev. Lett. {\bf 86} 1384 (2001); Phys. Rev. D {\bf 64},
084003 (2001).
\par
6. L. Lindblom and B.J. Owen, Phys. Rev. D {\bf 65} 063006 (2002)
\par
7. P. Haensel, K.P. Levenfish and D.G. Yakovlev, Astron Astrophys.
{\bf 381} 1080 (2002)
\par
\par
8. R.V. Wagoner, {\em Conditions for Steady Gravitational Radiation from
Accreting Neutron Stars} \htmladdnormallink{astro-ph/0207589}{http://arxiv.org/abs/astro-ph/0207589}
\par
9. N. Andersson, D.I. Jones and K.D. Kokkotas, {\em  Strange stars as
persistent sources of gravitational waves}
preprint \htmladdnormallink{astro-ph/0111582}{http://arxiv.org/abs/astro-ph/0111582}

\par
10. J. Heyl, {\em LMXBs may be important LIGO sources after all}
\htmladdnormallink{astro-ph/0206174}{http://arxiv.org/abs/astro-ph/0206174}
\par
11. For references to other potentially fatal mechanisms including magnetic
field wind-up, and turbulent viscosity and enhanced shear viscosity in a
boundary layer near the crust, see for example, the review Int.J.Mod.Phys.
D10 381-442,(2001), by N. Andersson and K.D. Kokkotas.

\vfill
\pagebreak
\section*{\centerline {
Gravitational waves from bumpy neutron stars}}
\addcontentsline{toc}{subsubsection}{\it  
Gravitational waves from bumpy neutron stars, by Ben Owen}
\begin{center}
Benjamin J. Owen, Pennsylvania State University
\end{center}

It has often been overlooked in the recent fuss about unstable oscillation modes,
but there is another way of getting gravitational waves out of single neutron
stars. You might even say a more solid way: A crystalline crust might not be
symmetric about the rotation axis of a star, giving the star a time-varying
quadrupole moment.

This lack of symmetry could come either in the form of localized bumps
(mountains), or a ``bump'' covering most of the star if the crust is symmetric
about an axis different from the rotation axis. One problem is that, due to the
nuclei having such low charge-to-mass ratios (being neutron-rich isotopes) and
being so tightly packed, the so-called solid crust has essentially the same
mechanical properties as Jell-O: It doesn't support bumps very well. Worse than
Jell-O, neutron stars are expected to support bumps no higher than $10^{-7}$
times the star's radius. (For the derivation of this number, and other details
and background, see the recent review by Ian Jones~[1].)

Numerous mechanisms have been proposed since the 1970s for producing bumps
or misaligned axes including strong ($10^{16}$G) magnetic fields, {\it Magnus
forces} due to superfluid vortex pinning, and just plain settling as an aging
star spins down and loses its centrifugal bulge. Until recently, however, there
wasn't much hope of detecting gravitational waves from any bumpy neutron stars
with LIGO or a comparable interferometer~[2].

The revival of interest in bumpy neutron stars as gravitational-wave sources
started about the same time as for unstable modes, when in 1998 Lars
Bildsten~[3] suggested that {\it electron capture} could make large mountains
on the accreting neutron stars in low-mass x-ray binaries. The density of a
neutron star crust doesn't increase smoothly as you go down towards the core,
but rather in discrete jumps. The result is layered like an onion: At the
bottom of each layer, a proton in each nucleus captures an electron due to the
intense pressure and turns into a neutron (inverse beta decay). This changes
the chemical composition in the next layer and allows the now less positively
charged nuclei to come closer together. Since the pressure of each electron
capture falls strongly with rising temperature, a rain of hot accreting matter
falling unevenly around the star can create buried mountains many layers deep.

Bildsten revived an old argument by Wagoner~[4] to show that, if there's a
bump, you can work out the gravitational wave strain from the observed x-ray
flux---assuming that torque-up from accretion is balanced by torque-down from
gravitational waves. Some rapidly-accreting x-ray binaries, particularly Sco
X-1, would produce gravitational wave strains as high as a few times $10^{-26}$
provided the assumption of torque balance holds.

Brady and Creighton~[5] examined the details of data analysis and detection.
Sco X-1 would be quite detectable by advanced LIGO (or a comparable
interferometer) if its orbit and the small fluctuations of its spin about the
torque-balanced value were precisely known, e.g.\ by radio observations. Sco
X-1 and most of its cousins are seen only in x-rays, where the timing and
orbital data are less precise. The need to search a large parameter space of
possible data demodulations increases the minimum observable strain by 2.5,
meaning that a broadband advanced LIGO configuration could barely detect Sco
X-1 above the noise. A narrow-band signal-recycling configuration such as
prototyped by GEO600 could get a signal-to-noise ratio several times higher. A
good detection above the noise would allow the extraction of interesting
physics of the crust such as breaking strain and probable thickness, which in
turn tells us something about the equation of state.

Is torque balance a reasonable assumption? Ushomirsky {\it et al.}~[6] checked
by combining nuclear physics with geology.  The bad news is that mountains
settle into the mantle as well as poke up into the sky, reducing the
quadrupole. The good news is that on an accreting neutron star the mountains
extend over many layers, which brings the quadrupole back up. If a star has a
5\% temperature variation over the surface and if the breaking strain of the
crust is $10^{-2}$, even the fastest-accreting binaries such as Sco X-1 can
achieve torque balance. More good news is that the crust won't smooth out a 5\%
temperature variation by conduction if the accretion is that irregular (and
constant). More bad news is that $10^{-2}$ is on the very high end of predicted
breaking strains, much higher than for terrestrial materials.

Another way of getting gravitational waves out of even axisymmetric neutron
stars is {\it free precession.} If the crust's symmetry axis is tilted away
from the rotation axis, both will precess about the fixed angular momentum
vector. This free precession will radiate gravitational waves and tend to get
damped both by radiation and by internal dissipation---the latter fairly
quickly---and so the topic has been pretty quiet since the 1970s.

Free precession got a lot more notice recently when Stairs {\it et al.}~[7]
discovered a pulsar that shows the expected modulation in its radio signal. The
star spins far too slowly for gravitational radiation to be significant, but it
shed some light on the physics of precession (if there are vortices in a
superfluid core they can't be pinned to the crust). Jones and Andersson~[8]
then systematically inventoried all known pumping mechanisms that could
encourage rapid free precession and strong gravitational waves. Unfortunately
the resulting signals are too faint even for advanced interferometers. Cutler
and Jones~[9] got excited briefly by the prospect of precessing neutron stars
being unstable to gravitational radiation like the $r$-modes, but on closer
inspection it turns out that an old result was wrong and radiation reaction
always damps precession.

Stairs {\it et al.}\ aren't the only ones with a detection paper. Middleditch
{\it et al.}~[10] also claim to have observed a freely precessing pulsar with a
2ms spin period in the remnant of supernova 1987A. They also claim that its
period derivative is consistent with spindown due to gravitational radiation!
Obviously this is really hot if true, but there are some problems. Like the
electron-capture scenario for Sco X-1, Middleditch's SN1987A would need a
breaking strain of $10^{-2}$---tough but conceivable. Furthermore, the extra
moment of inertia due to the bump would have to have changed by a factor of 2
in a few years while somehow maintaining a constant precession angle, which is
much harder to believe. Finally, the pulsar has been seen only by Middleditch's
team, and then only sporadically in the 1990s despite close observation.

Very recently the old mechanism of bumps through strong magnetic fields got a
new twist from Curt Cutler~[11]. It has been rediscovered several times since
the 1970s that a magnetic field of more than a few times
$10^{12}$G---a reasonable value for some young neutron stars---is secularly
unstable in a neutron star with an elastic crust. That is, if the magnetic
field axis is misaligned with the star's rotation axis, flexing of the crust
due to free precession will let the field move into its lowest energy
configuration. This happens to be one where the field axis is perpendicular to
the rotation axis. Cutler points out that an internal toroidal field can be
made quite strong by differential rotation in a young neutron star while
keeping the external dipole magnetic field low, consistent with observations.
This also means that gravitational radiation braking dominates electromagnetic
braking. As a result stars in various scenarios---x-ray binaries, recycled
millisecond pulsars, even (briefly) newborn neutron stars---could then be
detectable by advanced interferometers.

To sum it up: The goods may be odd, but the odds are pretty good that within a
few years we'll be seeing gravitational-wave signals from some known pulsars
with big bumps on them. The old bumping mechanisms from the 1970s have been
pepped up considerably, and with the observation of at least one precessing
pulsar hopes are high for more to come.

{\bf References}

[1] D. I. Jones, Class.\ Quantum Grav.\ {\bf 19}, 1255 (2002). Proceedings of
the 4th Amaldi Conference.

[2] P. R. Brady {\it et al.}, Phys.\ Rev.\ D {\bf 57}, 2101 (1998).

[3] L. Bildsten, Astrophys.\ J. {\bf 501}, L89 (1998).

[4] R. V. Wagoner, Astrophys.\ J. {\bf 278}, 345 (1984).

[5] P. R. Brady and T. Creighton, Phys.\ Rev.\ D {\bf 61}, 082001 (2000).

[6] G. Ushomirsky, C. Cutler, and L. Bildsten, MNRAS {\bf 319}, 902 (2000).

[7] I. Stairs, A. G. Lyne, and S. L. Shemar, Nature {\bf 406}, 484 (2000).

[8] D. I. Jones and N. Andersson, MNRAS {\bf 331}, 203 (2002).

[9] C. Cutler and D. I. Jones, Phys.\ Rev.\ D {\bf 63}, 024002 (2000).

[10] J. Middleditch {\it et al.}, New Astronomy {\bf 5}, 243 (2000).

[11] C. Cutler, \htmladdnormallink{astro-ph/0206051}{http://arxiv.org/abs/astro-ph/0206051}.

\vfill
\pagebreak
\section*{\centerline {
LIGO science operations begin!}}
\addcontentsline{toc}{subsubsection}{\it  
LIGO science operations begin!, by Gary Sanders}
\begin{center}
Gary Sanders, Caltech
\htmladdnormallink{sanders@ligo.caltech.edu}
{mailto:sanders@ligo.caltech.edu}
\end{center}

At 8:00 am Pacific time on August 23, LIGO began its first science
run, dubbed S1. As the S1 run draws to a close on September 9 it
appears to have succeeded in delivering a large data set from which
the first analysis effort for scientific results now departs.

In the previous report in MOG, Stan Whitcomb described the dress
rehearsal for S1, the engineering run E7. Since E7, the LIGO
Scientific Collaboration (LSC) upper limits analysis groups have
labored over the data, exercising the analysis codes, developing
algorithms and characterizing instrumental properties. The
interferometer commissioning teams, consisting of personnel from the
LIGO Laboratory and the LSC, have learned much from E7 and
incorporated the insights into further development of the
interferometers in the march to design sensitivity and to sustained
predictable detector operation.
\begin{figure}[h]
\centerline{\psfig{file=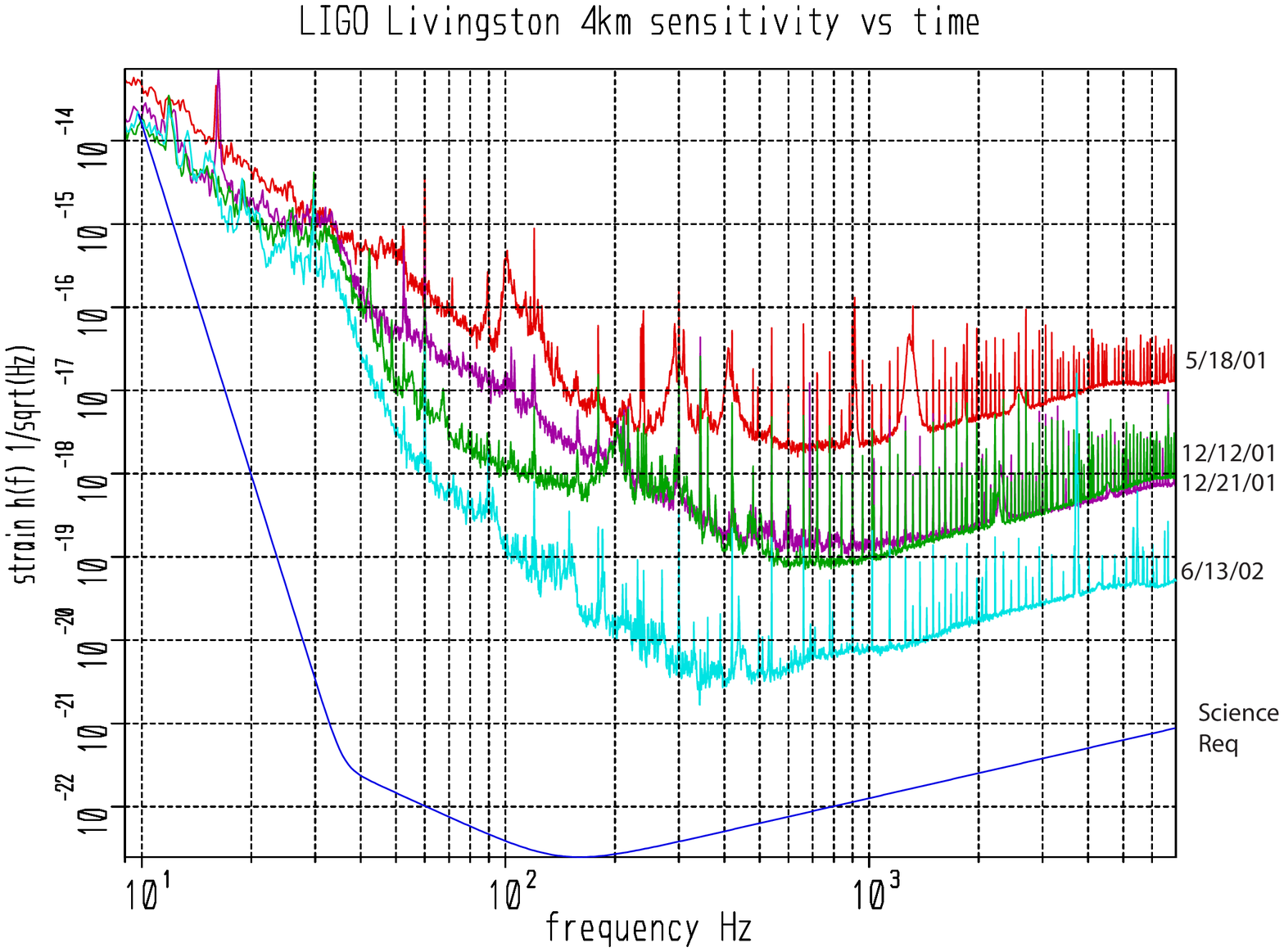,height=15cm}}
\end{figure}
To benchmark the progress in interferometer sensitivity, Figure 1
displays the progress in approaching LIGO design sensitivity for the
Livingston 4 km interferometer. The December 2001 curves mark the
sensitivity before E7. As LIGO began S1, the three LIGO
interferometers were much improved and roughly equally matched in
sensitivity and this is displayed in Figure 2.
\begin{figure}[h]
\centerline{\psfig{file=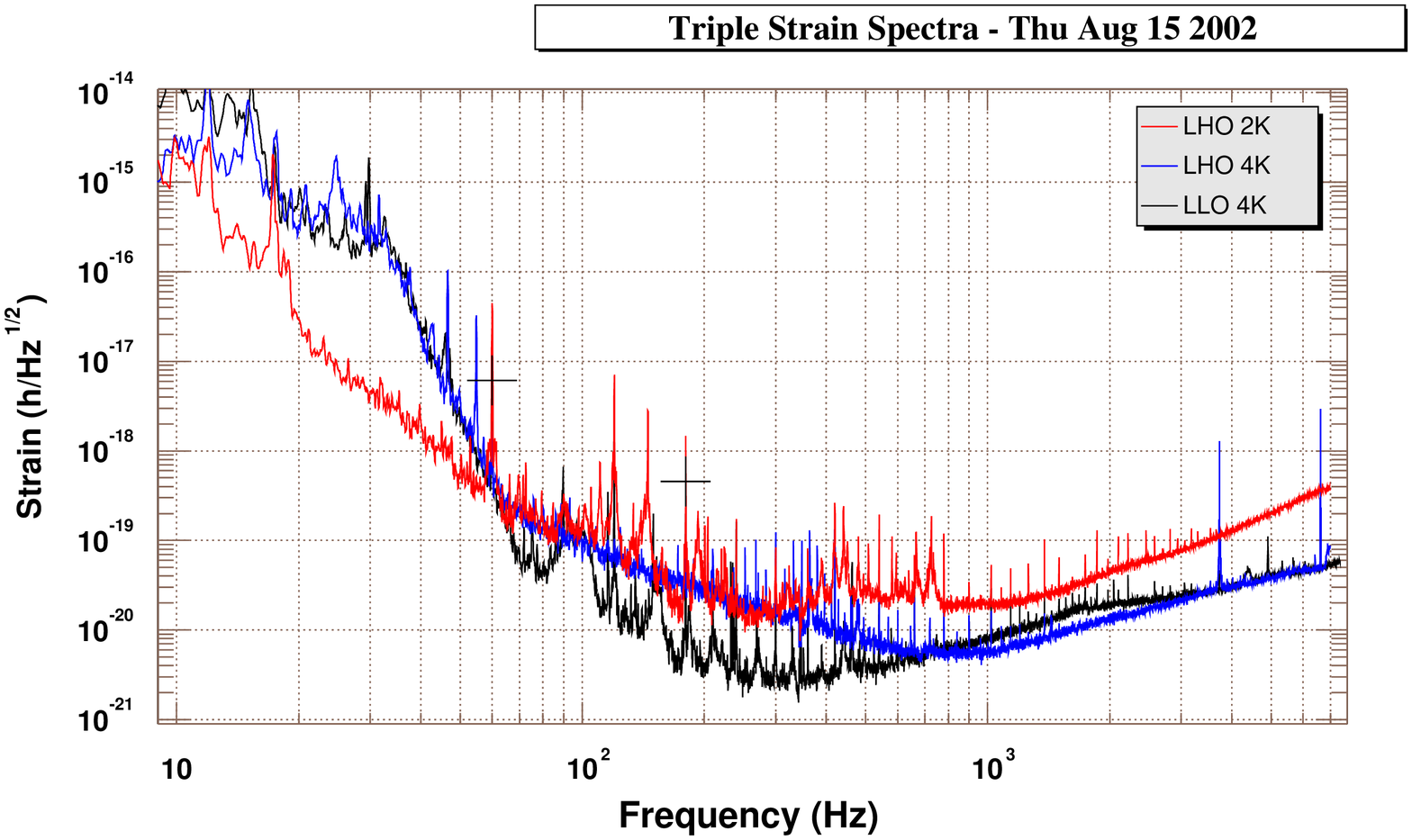,height=8.5cm}}
\end{figure}

All three interferometers have operated in the power-recycled mode
with individual interferometers operating in good "science"
configuration roughly between 50\% and 70\% of the time. Intersite
coincidences in lock have been logged typically about 1/3 of the
time. Triple coincidences are logged in the data about 20 - 25\% of the
time.

As in E7, LIGO and GEO have coordinated their running and the GEO 600
interferometer has been operating with high reliability much of the
time. LIGO and GEO have also been joined in coincidence with the
veteran TAMA 300 interferometer which skirted construction
disturbances at the TAMA site by running in coincidence with LIGO and
GEO over the August 30 to September 2 weekend. Arrangements have been
put into place between LIGO and GEO, and LIGO and TAMA for coordinated
analyses of the data sets.

The LIGO Scientific Collaboration is planning to complete a scientific
analysis of the S1 data by the end of the year, prior to S2, the next
in the series of progressively more sensitive science runs. S2 and S3
should be carried out at much improved sensitivity, and these runs
will extend for progressively longer periods.,

LIGO is funded by the US National Science Foundation under Cooperative
Agreement PHY-0107417. This work is a collaborative effort of the
Laser Interferometer Gravitational-wave Observatory and the
institutions of the LIGO Scientific Collaboration.

More information about LIGO can be found at: 
\htmladdnormallink{http://www.ligo.caltech.edu}
{http://www.ligo.caltech.edu}

\vfill\eject
\section*{\centerline {Fourth international LISA symposium}}
\addtocontents{toc}{\protect\medskip}
\addtocontents{toc}{\bf Conference reports:}
\addcontentsline{toc}{subsubsection}{\it  
Fourth international LISA symposium, by Peter Bender}
\begin{center}
Peter Bender, JILA, University of Colorado
\htmladdnormallink{pbender@jila.colorado.edu}
{mailto:pbender@jila.colorado.edu}
\end{center}
\parskip=3pt
\vspace{-0.3cm}
The 4th International LISA (Laser
Interferometer Space Antenna) Symposium was held at Penn State on July
20-24, 2002.  It followed earlier Symposia at the Rutherford Appleton
Laboratory near Oxford (1996), at Caltech (1998), and at the Albert Einstein
Institute in Golm, Germany (2000).

The present status of LISA is that it is being planned as a joint mission of
ESA and NASA.  It officially is a Cornerstone Mission in the ESA program,
but with the time schedule to be determined in consultation with NASA.  From
the NASA side, LISA is in the 5 Year Strategic Plan of the Office of Space
Science, which means that the OSS hopes to obtain approval for the mission
within 5 years.  For planning purposes, a possible launch date of about 2011
is being considered.

The Symposium had 9 half-day sessions.  The first gave a broad view of
gravitational wave astronomy in the LISA epoch, and started with a talk on
Science with LISA by Bernard Schutz.  Next, Karsten Danzmann described LISA
Technology and the Future.  The last two talks were by David Shoemaker on
Ground-Based GW Interferometers in the LISA Epoch, and by Massimo Cerdonio
on Acoustic GW Detectors for the 2012 Timeframe.

Three of the remaining sessions covered LISA Sources and Source Science.
One talk by Gijs Nelemans discussed galactic binaries, including their
abundance and what can be learned from their signals.  The main topic,
however, was massive black holes, and their important role in astrophysics.

One important issue is how black holes grow to the roughly 
$10^5 M_{\rm Sun}$ or
larger range, or whether they form with high masses by sudden collapse of
large gas clouds or supermassive stars.  Another issue is whether information
about galaxy formation can be obtained from coalescences of massive black
holes already present in pregalactic or galactic structures that merge.
Existing and new information relevant to these topics was given in a number
of talks by astrophysicists, that were among the highlights of the
Symposium. As an example, Tom Abel discussed calculations of the probable
mass distribution of the first stars to form, with the mass range up to
about $300 M_{\rm Sun}$ appearing to have been favored.  
Thus many roughly $100 M_{\rm Sun}$
black holes may have formed early from rapid stellar evolution, and some
could have gotten into galactic nuclei where massive black holes were being
formed.          

Another topic discussed in some detail was coalescences of stellar mass
black holes and compact stars with massive black holes.  Such highly unequal
mass coalescences occur when compact objects in the density cusp around a
massive black hole get scattered in close enough to start losing substantial
energy by gravitational radiation.  Sterl Phinney and Steinn Sigurdsson
described both recent and earlier work in this area.  In addition, the
strong tests of general relativity that can be obtained from observing such
events or more equal mass coalescences were discussed by Cliff Will.

Some of the other theoretical and data analysis topics included modeling
binary black hole coalescence, radiation reaction in strong fields,
formation of neutron stars and stellar mass black holes, templates for
highly unequal mass coalescences, analysis of galactic binary signals, and
the higher harmonic GW content of black hole binary inspiral signals.  Tom
Prince finished up the sessions on sources by reviewing the LISA science
requirements and science challenges.

The highlights of the experimental and technology development sessions were
the detailed discussions of the gravitational reference sensors, which also
can be described as free mass units.  Each of these devices provides a
freely floating test mass inside a housing with capacitive electrodes on its
inside to detect any relative motion.  What gives LISA its high sensitivity,
even at frequencies below $10^{-4} Hz$, 
is keeping the spurious accelerations of
the test masses extremely low.

The three LISA spacecraft will be in solar orbit roughly 20 deg. behind the
Earth, and form a nearly equilateral triangle 5 million km on a side.
Micronewton thrusters are used to keep the spacecraft nearly motionless with
respect to the test masses, despite solar radiation pressure and other
non-gravitational forces on them.  Laser distance measurements are made
continuously between the test masses in the different spacecraft, and
extremely small GW changes in the distances at periods of roughly a day to a
second can be detected.

Almost all aspects of the mission can be tested on the ground.  However, it
does not appear possible to ground-test the gravitational reference sensors
for spurious accelerations to the desired level.  Thus ESA is planning a
technology validation mission called SMART-2 for a 2006 launch, with
demonstrating the gravitational reference sensor performance and the
micronewton thrusters as the main objective of the mission.  A European LISA
Technology Package will be carried to accomplish this, as well as a
NASA-provided package called the Disturbance Reduction System.  The decision
was made recently to provide the US package under the NASA New Millennium
Technology Validation Program, since the principles demonstrated will be
valuable for a number of other US missions besides LISA.

Two of the major talks were by Stefano Vitale and by William Folkner on the
ESA and NASA test packages respectively.  They laid the groundwork for
additional talks on the detailed design of the units, and on the analysis of
the noise sources for them.  These included discussions of interferometric
measurements of the tiny residual motions of the test masses, control of
charging of the test masses due to cosmic rays, mechanisms for caging the
test masses during launch, flight experience with accelerometers having many
similarities to the gravitational reference sensors, and overall modeling of
the expected performance.

Other LISA experimental and technology development talks included two by
John Armstrong and Jean-Yver Vinet on the ``time-delay interferometry"
approach that will be used in LISA.  This method has been developed for
cancelling out the effects of phase noise in the lasers and some other
undesired effects.  Plans for making the required phase measurements on the
laser signals also were described, and European work on development of the
micronewton thrusters was reviewed.  In addition, reports were given on the
modeling of the motions of the coupled spacecraft and test mass systems, and
on modeling of the whole LISA mission.

One other session was included to review progress on the design and
construction of ground-based GW detectors.  This included interferometric
detectors such as LIGO, VIRGO, GEO-600, and TAMA, and also acoustic
detectors.  TAMA already is operating fairly frequently, and the first
science run for LIGO is scheduled for starting in August, with parallel
operations planned by GEO-600 and TAMA.  Progress also is continuing with
the acoustic detectors, including the development of the first spherical
detectors, MiniGRAIL and two others, in The Netherlands, Brazil, and Italy.

The  Symposium was sponsored by the National Science Foundation Center for
Gravitational Wave Physics at Penn State. Co-sponsors included NASA's Jet
Propulsion Laboratory, and travel assistance for some European attendees was
provided by the European Space Agency and the new Albert-Einstein-Institute
for Experimental Gravitation.  The very efficient and helpful Chair of the
Scientific Organizing Committee was Sam Finn from Penn State.

\vfill\eject
\section*{\centerline{Initial Data for Binary Systems}}
\addcontentsline{toc}{subsubsection}{\it  
Initial Data for Binary Systems, by Gregory Cook}
\begin{center}
Gregory B. Cook, Wake Forest University
\htmladdnormallink{cookgb@wfu.edu}
{mailto:cookgb@wfu.edu}
\end{center}
\parskip=5pt

The {\em Center for Gravitational Wave Physics} at the Pennsylvania
State University held its first {\em Focus Session} on March 29--30,
2002.  The topic was ``Initial Data for Binary Systems''.  Organized
by Pablo Laguna and myself, the meeting hosted eight invited speakers
and thirty participants.  The primary goal of the workshop was to
foster discussion on current open questions and possible future
directions related to the construction of astrophysically relevant
initial data representing binary systems containing black holes and
neutron stars.  This is a very difficult standing problem for the
LIGO/LISA source-modeling community.  There are currently a number of
different approaches that can be used to construct compact-binary
initial data.  Unfortunately, the only thing we are sure of is that,
in one way or another, all of these current methods fall short of the
goal of representing astrophysical systems with sufficient fidelity.
With these initial data serving as the starting point for full
numerical simulations of the plunge and coalescence of compact
binaries, it is clear that, at the very least, the quality of the
initial data needs to be well understood.

Why is it so hard to construct realistic initial data for compact
binary systems?  A combination of factors come into play.  The initial
value problem of general relativity is a {\em constrained} hyperbolic
system.  At any given instant of time, the gravitational fields must
satisfy four constraints that can be posed as a set of elliptic
boundary-value equations.  Solving these equations represents the
primary computational difficulty in constructing initial data, and
this aspect of the problem has received considerable attention.
However, solving the constraints only fixes one-third of the
degrees-of-freedom of the gravitational fields.  The remaining
degrees-of-freedom are divided evenly between the gauge freedom of the
theory and the freely specifiable initial dynamical content of the
gravitational fields.  Historically, the choices for these latter
degrees-of-freedom have been based largely on what would simplify the
problem of solving the constraints, not on what would produce the most
realistic data.  There are of course additional problems.  The exact
definition of {\em astrophysically realistic} initial data is not
fully understood.  Furthermore, given only the initial data for a
gravitational system, it is impossible (except for special cases) to
determine its full physical content.  This requires evolving the data.

The schedule for the {\em Focus Session} was designed to foster active
discussion.  There were four sessions over two days.  Each session was
limited to two half-hour invited talks, with each talk followed by an
hour of discussion.  Participants were encouraged to prepare one
or two transparencies and to ask the session chairs for time to
present these at an appropriate time during the discussion sessions.
The first day offered talks by Peter Diener, Philippe Grandcl\'ement,
Richard Matzner and myself.

Philippe Grandcl\'ement and I each presented new approaches for
constructing binary black hole initial data.  These approaches are
very similar and try to extend to black holes an approach that has
been very fruitful for the case of neutron stars.  They incorporate an
approximate helical Killing vector, or a notion of quasi-equilibrium,
into the process of constructing the data.  The most important feature
of these approaches is that they employ a much different method for
fixing the extrinsic curvature.  For black holes, essentially all
approaches for constructing initial data have used an analytic
solution of the momentum constraints (the Bowen-York solution) to fix
the extrinsic curvature.  Recent attempts to improve black hole data
have used superpositions of the extrinsic curvature of a boosted Kerr
black hole as a foundation for the extrinsic curvature.  A difficulty
with either of these approaches is that they incorporate an unknown
contribution to the freely specifiable dynamical content of the data.
Loosely speaking, some amount of unphysical junk is built into the
data.  In the new approaches discussed, the only assumption built into
the extrinsic curvature is that it should, at least instantaneously,
lead to a stationary geometry.  (Technically only the conformal
three-geometry is instantaneously stationary.)  This new approach for
specifying the initial data for black-hole binaries in quasi-circular
orbits seems to yield results that are in better agreement with
post-Newtonian results.

Richard Matzner presented a talk on using a superposition of boosted
Kerr geometries as background data for solving the constraints. He
also discussed many issues related to the physics of close binaries:
the meaning of the innermost stable circular orbit (ISCO), the effects
of spin and frame dragging, and tidal forces.  This use of superposed
boosted Kerr geometries as background data for solving the constraint
equations is among the first attempts to employ more realistic choices
for the freely specifiable initial dynamical content of the initial
data.

Another theme of the meeting was to explore the possibility of
incorporating, into initial data, information from post-Newtonian
solutions for binaries in circular orbits.  Peter Diener presented
what may be the first attempt to use this information in fixing the
background fields for both the metric and extrinsic curvature.  This
approach relies on using post-Newtonian solutions in the ``ADM
transverse-traceless'' gauge.  It seems that a major difficulty with
the idea of incorporating post-Newtonian information is that different
parts of the post-Newtonian solution are only well defined in certain
regions (near-zone, wave-zone, far-zone).

These issues were explored again on the second day of the {\em Focus
Session} when talks were given by Bala Iyer, Thomas Baumgarte, K\=oji
Ury\=u, and Olivier Sarbach.  Bala Iyer gave an overview of how
waveforms for inspiralling compact binaries are computed in
post-Newtonian formalisms.  This led to an extensive discussion about
what information from the various post-Newtonian calculations could be
incorporated into a background metric for use in solving the
constraints.  The consensus was that much of the wave information
cannot be taken directly from current post-Newtonian calculations.
One suggestion, however, was to use some kind of numerical
post-Minkowski approach to obtain wave information that could be
incorporated into a background metric.

Thomas Baumgarte and K\=oji Ury\=u each presented approaches that used
sequences of individual initial-data sets to model the inspiral of a
neutron-star binary system.  While their approaches were rather
different, both made the assumption of a quasi-adiabatic inspiral and
computed the gravitational waveforms being produced.  In Thomas
Baumgarte's approach, each initial-data configuration was evolved in a
full dynamical code, with the restriction that the neutron-star matter
sources were ``frozen'' in the corotating frame.  After several
orbits, the configuration reached a steady state and the gravitational
waves being emitted were extracted.  K\=oji Ury\=u's approach differed
in that it solved for a perturbation of the initial metric, evolved a
linearized system, and expanded the perturbation in spherical
harmonics.  Although they followed different approaches, they both
computed the gravitational-wave luminosity and then estimated the
radiation reaction timescale in order to produce waveforms.  These
approaches seem to be promising avenues for computing the
gravitational waveforms down to a point near the ISCO, and for
providing corrections to the background metric which can be used to
improve initial-data computations.

In the final talk, Olivier Sarbach presented results for a rather new
approach for constructing black hole initial data.  This approach
differs from others in that it constructs initial data that is
strictly of the Kerr-Schild type.  While it isn't clear if this
restriction will allow for constructing astrophysically realistic
binary initial data, it is well suited for constructing data that can
be used in ``close-limit'' perturbative evolutions.  Such computations
are especially useful for comparison with full nonlinear evolution
codes.

I want to close by offering my thanks to the directors and staff of
the {\em Center for Gravitational Wave Physics} for their support in
running this {\em Focus Session}.  They did a great job.  Especially
enjoyable was a wonderful banquet, hosted by Abbay Ashtekar, held
Friday evening at a local Indian restaurant.  A great time was had by
all.

A full listing of the talks, along with copies of the speakers'
slides, can be found at\\ 
\htmladdnormallink{http://cgwp.gravity.psu.edu/events/InitialData}
{http://cgwp.gravity.psu.edu/events/InitialData}.

\vfill\eject
\section*{\centerline {Joint LSC/Source Modeling Meeting,  20-24
March 2002}}
\addcontentsline{toc}{subsubsection}{\it
Report on Joint LSC/Source Modeling Meeting, by Patrick Brady}
\begin{center}
Patrick R Brady, University of Wisconsin-Milwaukee\\
\mbox{\ }
\htmladdnormallink{patrick@gravity.phys.uwm.edu}
{mailto:patrick@gravity.phys.uwm.edu}
\end{center}

In parallel with the  March meeting of the LIGO Scientific
Collaboration (LSC), there was a workshop to discuss formation of
source working groups for interferometric gravitational-wave
detectors.  The workshop was motivated by the growing need for a
tighter coupling of the gravitational-wave source-analysis community
to the experimental gravitational-wave projects.  The meeting took
place at the LIGO Observatory in Livingston, Louisiana over three
days interlaced between the LSC meeting.   Many of the presentations
are available from the LSC web site [1].

On the evening of the first day,  the source analysts gave a series of
short (15 minute) presentations about the state of source modeling as
a field.   Over the course of a long evening from 4pm until 10pm,
attendees were shown the breadth of ongoing work.   The
presentations covered inspiral of compact binaries,  binary black hole
mergers, binary neutron star mergers,  neutron-star black-hole
mergers, stellar collapse and neutron star vibrations.   Despite the
short time available for each presentation and the long evening,
presenters attempted to give a flavor of their work,  where it stands,
and where it is going.  

The second day of the meeting was given over to the formal sessions of
the LSC in the morning,  and to presentations about how source analysts 
might interface with data analysts in the afternoon.

The third day of the meeting began with a series of
presentations/panel discussions by data analysts.   These
presentations were intended to initiate the discourse on the need for
information about sources.   Considerable emphasis was placed on data
analysis techniques other than matched-filtering,  which the source
analysis community is already familiar with.    The discussion focused
on the use of time-frequency methods in the detection of
gravitational waves,  emphasizing how information from the source
analysts can be used to develop data analysis algorithms even in the
absence of knowing the full waveform.  Two examples of urgent needs
for the ground-based detectors are:   (1) Information on the phase
evolution of waves from the late (``IBBH") stage of compact binary
inspiral (where post-Newtonian expansions fail).  (2)  Information
from numerical relativity simulations of black-hole merger waveforms
which can be used in applying the excess-power algorithm.  During the
session on LISA data analysis,   the needs were shown to be somewhat
different at this time.   For example,  the first urgent need is to
acquire sufficient science information to firm up LISA's noise curve
by December 2003.

By the end of the session,  it was clear that source analysis should
be guided by the data analysis needs,  and the development of data
analysis algorithms will be more effective if source analysts
participate to some degree.   As the community moves from the goal of
first detection of gravitational waves toward gravitational-wave
astronomy,  interpretation of the data will inevitably lead to closer
coupling between source analysts and data analysts;  when that
happens,  those with experience will benefit most from the mutual give
and take

Separate from this meeting,   NASA and NSF have been discussing a possible
new initiative to fund source modeling efforts in the gravitational
wave community.   Members of a joint NASA/NSF panel,
chaired by Saul Teukolsky,  were present at the meeting and took the
opportunity to obtain feedback for their report.   An open discussion,
in which Teukolsky explained the panel's mandate\footnote{The panel's mandate
was to identify the most urgent needs for LIGO and LISA in the area of
theory involving large-scale computing and,  if necessary,  to
identify the level of additional resources required to deal with those
needs.} took place on the afternoon of the second day.   The
discussion was animated and helped to identify the needs of the
community.   The report from the Teukolsky panel is now 
available [2].

The third day finished with a discussion of how to coordinate the
efforts of the two communities more coherently.   Several different
methods were advocated:  closer connection with the experiments [for
example, through the Astrophysical Source Identification and
Signatures (ASIS) group of the LSC], groups organized by source and
led by their members.  Once again,  the discussion was animated and
brought out problems with each of the possible ways to organize.   In
many cases,  the purpose of such organization was called into
question.  In the end, however,  it was clear that those involved in
target-source data analysis had a strong desire to bring the source
analysis community closer to the data analysis activities.  Moreover,
there is a community which wants a closer connection to the
gravitational-wave experiments without direct obligation to provide
deliverables to the LSC or another collaboration.   The outcome of the
discussions was the formation of several source groups to facilitate
communication among its members with two goals: 
\begin{enumerate} 
\item
The data analysts should educate the source analysts about
gravitational-wave data analysis, and most especially about what kinds
of source-analysis information will be useful, and in what ways, in
data analysis.
\item The source analysts should educate the data analysts about
source analysis and simulations, and most especially about what kinds
of information it will be possible to supply for the data analysis and
on what timescales.
\end{enumerate}

Following the workshop in March,  five source groups have formed.  (In
fact, another group centered on sources of stochastic gravitational
waves is also forming.)   At least two facilitators have been
identified within each group (one data analyst and one source analyst)
in the hope that they can help the groups get started.   The
facilitators' roles may become less important as the groups begin
to self-organize.   Information about the groups can be found at 
\htmladdnormallink{http://www.lsc-group.phys.uwm.edu/gwawg}
{http://www.lsc-group.phys.uwm.edu/gwawg} along with links about
membership and to mailing lists for each group.   Anybody interested
in participating should join the mailing list which will be the venue
for preliminary discussion/organization.   The five existing groups
are listed here with a descriptive paragraph adapted from the web site:
\begin{description}
\item[Inspiral of Comparable Mass Binaries] 
       This group will focus on the late stages of inspiral, up to the
       final plunge and merger,  for BH/BH, NS/BH, and NS/NS binaries.
       Source analysis is currently being carried out via
       post-Newtonian methods, resummation methods,  and numerical
       relativity methods.  Data analysis is currently via matched
       filters using post-Newtonian (or kludged) waveforms, and by
       the fast chirp transform.
       
       \emph{Source Analysis Facilitators}: Bala Iyer and Thomas Baumgarte\newline
       \emph{Data Analysis Facilitator}: Jolien Creighton

\item[Binary Black Hole Mergers]
       This group will focus on plunge from the innermost stable
       circular orbit and the final merger of binary black holes.
       Source analysis is currently by numerical relativity
       techniques.  Data analysis is currently via various
       time-frequency techniques.  
       
       \emph{Source Analysis Facilitators}: Bernd Bruegman and Luis Lehner\newline
       \emph{Data Analysis Facilitator}: Patrick Brady

\item[NS/NS and NS/BH Binary Mergers] 
       This group will focus on plunge induced by combined general
       relativity and tidal couplings,  bar formation and evolution,
       oscillations of neutron stars. It will also focus on tidal
       disruption of the neutron star by the black hole in NS/BH
       binaries.  Source analysis is currently by numerical
       simulations (Newtonian, post-Newtonian, and fully
       relativistic).  Data analysis is currently via various
       time-frequency techniques.
       
       \emph{Source Analysis Facilitators}: Fred Rasio and Masaru Shibata\newline
       \emph{Data Analysis Facilitator}: Ben Owen

\item[Stellar Collapse]
      This group will focus on core collapse at end of stellar
      evolution to form a neutron star, a centrifugally hung-up
      protoneutron star, and/or a black hole; accretion-induced
      collapse of a white dwarf;  collapse of a supermassive star for
      LISA;   dynamical or secular instabilities of neutron stars.
      Source analysis is currently by numerical simulations
      (Newtonian, post-Newtonian, and fully relativistic), and by
      study of stability and dynamics of equilibria models.  Data analysis is
      currently via time-frequency techniques.
      
       \emph{Source Analysis Facilitator}:  Tony Mezzacapa\newline
       \emph{Data Analysis Facilitator}:  Warren Anderson

\item[Inspiral of Compact Objects into Supermassive Black Holes]
     This group is already rather far along in its organization and
     work, and is attached to LISA Working Group 1.   It focuses on
     white dwarfs, neutron stars, and stellar mass black holes
     captured into small-periastron orbits around a supermassive
     black hole in a galactic nucleus; evolution of the orbits under
     radiation reaction and via perturbations of other orbiting objects
     and via interaction with accretion disks; transition to plunge
     and capture.  Source analysis is via black-hole perturbation
     theory and N-body simulations for capture computations as input
     for estimating capture rates.  Data analysis is expected to be by
     hierarchical techniques that involve mixed coherent
     (matched-filter) and incoherent methods.  Data analysis work
     includes separating the strongest sources from the background of
     weaker inspiral waves and WD/WD binary waves.
     
       \emph{Source Analysis Facilitator}: Scott Hughes\newline
       \emph{Data Analysis Facilitator}:  Curt Cutler
\end{description}
Most of these groups are embryonic and all interested parties are
encouraged to join in and help define their activities.    

{\bf References:}

[1] LIGO Scientific Collaboration web site: \hfil\break
\htmladdnormallink{http://www.ligo.caltech.edu/LIGO${}_{}$web/lsc/lsc.html}
{http://www.ligo.caltech.edu/LIGO${}_{}$web/lsc/lsc.html}.

[2] \emph{Preliminary Report of the Task Group for an
NSF/NASA Computational Initiative on Gravitational Wave Science} is
available from 
\htmladdnormallink{http://gravity.phys.psu.edu/\~{}tggweb/} 
{http://gravity.phys.psu.edu/\~{}tggweb/} 
or
\htmladdnormallink{http://astrogravs.gsfc.nasa.gov}  
{http://astrogravs.gsfc.nasa.gov}.  

\vfill\eject
\section*{\centerline {
Greek Relativity Conference, NEB-X}}
\addcontentsline{toc}{subsubsection}{\it Greek Relativity
Conference, NEB-X by Kostas Kokkotas and Nick Stergioulas}
\begin{center}
Kostas Kokkotas and Nick Stergioulas, Aristotle University of
Thessaloniki, Greece
{mailto:kokkotas@astro.auth.gr, niksterg@astro.auth.gr}
\end{center}

The 10th Greek Relativity meeting has been hosted by the
Thessaloniki group in Kallithea, Chalkidiki, an idyllic sea resort
near Thessaloniki, from May 29 till June 2, 2002. The series of
meetings of the Greek relativity community, named NEB (New
Developments in Gravity), was initiated by Basilis Xanthopoulos in
1984. Since then, every second year Greek Relativists working
either in Greek universities or abroad meet to present their
research work and to discuss recent developments in gravity.

This year's meeting
(\htmladdnormallink{http://www.astro.auth.gr/gravity/}
{http://www.astro.auth.gr/gravity/})attracted increased
international participation with respect to the previous ones, in
an attempt to promote NEB into a regional meeting for South-East
Europe. Several of the world's top relativists were invited and
delivered plenary talks in Cosmology, Mathematical Relativity,
Relativistic Astrophysics, Gravitational-Wave Detection and
Quantum Gravity

In the first session, the focus was on Cosmology and Brane
Gravity. Roy Maartens gave a review talk on the geometry, dynamics
and perturbations of brane-world models. The simplest of these
models are able to reproduce the predictions of general relativity
at low energies, while introducing interesting new features at
high energies - for example in the very early universe, or during
gravitational collapse to a black hole, or in cosmic ray showers.
A new method for constructing branes of any dimensionality was
presented by Nikolaos Batakis while Georgios Kofinas showed an
analysis of the induced brane dynamics, when the intrinsic
curvature term is included in the bulk action. The invariant
description of Bianchi-homogeneous 3-spaces, by considering the
action of the automorphism group in the configuration space of
real, symmetric and positive definite $3 \times 3$ matrices, was
the subject of the talk of Theodosios Christodoulakis. The session
continued with Nicolaos Spyrou, who presented work on the
conformal-invariance approach of hydrodynamical flows in
cosmological models. Leandros Perivolaropoulos  in his talk showed
that the redshift of pressureless matter density due to the
expansion of the universe, generically induces small oscillations
in the stabilized radius of extra dimensions (the radion field);
low-frequency oscillations lead to oscillations in the expansion
rate of the universe, which could naturally resolve the
coincidence problem. A dynamical systems approach to scalar field
cosmologies has been presented by John Miritzis, who analyzed
general mathematical properties of the differential equations
describing the evolution of FRW models. Argyris Nicolaidis
described how experiments may yield proof of the existence of
extra dimensions of space, and Christos Eleftheriadis concluded
the session with a presentation of the CAST experiment at CERN,
aiming at the detection of axions.

The second session, devoted to Mathematical Relativity, opened
with Jiri Bicak's  in-depth review of exact models of radiative
spacetimes.  Sayan Kar argued that the Kalb-Raymond field, coupled
gravitationally to the Maxwell field, can lead to a
wavelength-independent optical activity in synchrotron radiation
from cosmologically distant radio sources. Work aimed at finding
interior, anisotropic fluid solutions, matched to the Kerr metric,
was reported by Taxiarchis Papakostas, while the conditions under
which the nonradial stresses might prevent spherical gravitational
collapse were discussed by Petros Florides. Dimitris Tsoubelis
presented a new family of integrable nonlinear systems that
include the Ernst equation for colliding gravitational waves.
Anastasios Tongas discussed geometrical aspects of integrable
nonlinear equations of the Schwarzian type. Spyros Cotsakis
reported on work with Y. Choquet-Bruhat in which they prove
completeness theorems in general relativity under generic
geometrical assumptions, while George Papadopoulos presented a
talk on an alterative proof of the generality of the
Kantowski-Sachs vacuum, based on general coordinate
transformations that preserve spatial homogeneity. The definition
of conserved quantities in general theories of gravity was
discussed by Andreas Zoupas. The session ended with a presentation
by Christos Tsagas on how gravitational waves can produce and
sustain large-scale magnetic fields, strong enough to seed the
galactic dynamo.

The next session was devoted to Astrophysical Relativity and the
Detection of gravitational waves. Bernard Schutz gave a thorough
introduction on the most promising astrophysical sources of
gravitational waves for the new-generation detectors, while on the
experimental side Gabriela Gonzalez updated us on the current
status of the LIGO detectors. In the same session, Remo Ruffini
gave a historical review of the gamma-ray burst puzzle and
summarized the current theoretical understanding. 3-D numerical
simulations of rotating relativistic stars and the first
computation of their quasi-radial modes of pulsation in rapid
rotation were shown by Nick Stergioulas, followed by a report by
Theocharis Apostolatos on computations of differentially rotating
relativistic stars and their pulsations. The effect of
quasi-normal mode excitation on the detection of gravitational
waves from neutron star binaries was discussed by Emanuele Berti.
Kostas Kokkotas argued in his talk that, due to the $r$-mode
instability, strange stars can be a good and persistent source of
gravitational waves. Uli Sperhake talked about his studies on
nonlinear radial oscillations of relativistic stars viewed as
deviations from an equilibrium state, and Adamantios Stavridis
reported on recent computations of $r$-modes in relativistic
stars, showing that discrete modes exist in most cases, due to the
coupling of axial and polar terms. Dimitrios Papadopoulos talked
on acceleration and cyclotron radiation induced by gravitational
waves,  Sotirios Bonanos ended this session with a description of
the capabilities of his ``Riemannian Geometry and Tensor
Calculus'' package for Mathematica, which is freely available at
\htmladdnormallink
{http://www.inp.demokritos.gr/\~{}sbonano/RGTC/RiemannTensorCalculus.html}
{http://www.inp.demokritos.gr/\~{}sbonano/RGTC/RiemannTensorCalculus.html}.

The conference concluded with a special review by Jorge Pullin on
the stability properties of various discrete versions of
Einstein's equations, and their use in the canonical quantization
of general relativity.

The next (11th) Greek Relativity meeting will take place in
Lesvos, in the summer of 2004
\htmladdnormallink{http://www.astro.auth.gr/gravity/NEB.html}
{http://www.astro.auth.gr/gravity/NEB.html}.

\vfill\eject
\section*{\centerline {
Gravity, Astrophysics, and Strings @ the Black Sea}}
\addcontentsline{toc}{subsubsection}{\it
Gravity, Astrophysics, and Strings @ the Black Sea, by Plamen Fiziev}
\begin{center}
Plamen Fiziev, Sofia University
\htmladdnormallink{fiziev@phys.uni-sofia.bg}
{mailto:fiziev@phys.uni-sofia.bg}

\end{center}

The first international conference Gravity, Astrophysics, and Strings
@ the Black Sea took place from June 10 till June 16, in a small town
on the Bulgarian sea shore, called Kiten. Somebody suggested that if
we rename the place by replacing the ``K'' with ``W'', the conference will
have much greater success. Actually, a brief account of the lectures
presented at the meeting, suggests that the endeavor achieved our main
goal - to bring together experienced researchers from the title
subjects with younger physicists and students, mostly from the South
Eastern corner of Europe. 
The meeting was organized by the Joint Group
in Gravity and Astrophysics, including researchers from various
institutions in Sofia. The participants were representing local
researchers, students, and scientists from major institutions in
roughly equal proportions. As the title suggests, our idea was to
gather researchers from relatively distant although related areas and
explore the consequences of this diversity.  

Edward Seidel (MPI)
opened the Conference by giving his first lecture on the Numerical
Relativity and Its Applications to Gravitational Wave Astronomy. In
total of 4 hour-long lectures the foundations of numerical relativity
ware described. The theoretical formulations of Einstein's equations,
tools and techniques for analyzing black hole space-times,
gravitational waves, boson stars, relativistic hydrodynamics, and
other topics were covered. Gabrielle Allen (MPI) introduced the CACTUS
Toolkit as a freely available, modular, portable and manageable
environment for collaboratively developing parallel, high-performance
multi-dimensional simulations. Jorge Pullin (Louisiana State U)
presented some new ideas on consistent discretizations in classical
and quantum gravity. A very interesting discussion on Supercomputing
in Europe completed the first day program. No question about it:
Indeed, there was a Welcome Party in the evening.  

On the second day,
D. Sasselov (Harvard U) acquainted the audience with the first atoms
in the Early Universe and the cosmic wave background radiation. Before
lunch, U. Sperhake (Aristotle U of Thessaloniki) presented a numerical
approach, which enables us to evolve radial oscillations of neutron
stars over large amplitude range with high accuracy. In the afternoon,
K. Kokkotas (Aristotle U of Thessaloniki) started his series of
lectures on the recent results on instabilities of different kind of
relativistic stars, and especially of neutron stars. Many new
numerical results of different scientific groups were reported. Also,
R. Rashkov (Sofia U) gave a review for students on strings, brane
worlds, and Ekpyrotic theory.  In these lectures a brief but intense
review was given of the progress in understanding String theory and
its implications to gravity, cosmology and gauge theories.  

The third
day started with some of the follow-ups of the abovementioned series
of lectures as well as a presentation by E. Guendelman (Ben Gurion U)
on two measure gravity theories and cosmology, in which the
possibility of using a measure of integration independent of the
metric was studied. In the afternoon a relaxing excursion took place
to the outfall of the nearby Ropotamo river. The remaining program of
the day included an account of spherically symmetric braneworld
solution given by E. Papantonopoulos (National Technical U of
Athens). After reviewing the known black hole solutions on the brane,
the induced gravity scenario was discussed, according to which an R(4)
term is induced on the brane. To close the scientific program for the
day, M. Vavoulidis (Aristotle U of Thessaloniki) gave a talk on
rotating relativistic stars.  

On the next day, right on time,
M. Bander (U of California, Irvine) resurrected time in quantum
gravity.  By allowing for non zero vacuum expectation values for some
of the fields that appear in the Hamiltonian constraint of canonical
general relativity a time variable, with usual properties, can be
identified; the constraint plays the role of the ordinary
Hamiltonian. Before lunch time, E. Horozov (Sofia U) talked about the
Weyl algebra and bispectral operators.  In the afternoon, another
entertaining event took place: The participants in the Conference
visited the old towns of Nessebar and Sozopol. These are towns,
settled for the first time by the ancient Greeks, with many churches
from Byzantine times and old fisherman houses. The closure for the day
was provided by the official dinner.  

On Friday, M. Todorov (Technical
U - Sofia) presented a joint work with T. Boyadjiev, P. Fiziev, and
S. Yazadjiev (Sofia U), on a new numerical algorithm for solving a
class of BVPs with internal free boundaries.  An investigation of
numerically models of the static spherically symmetric boson-fermion
stars in the scalar-tensor theory of gravity with massive dilaton
field was presented. The proper mathematical model of such stars is
interpreted as a nonlinear two-parametric eigenvalue problem with an
unknown internal boundary. After that, P. Fiziev (Sofia U) gave the
first of his pair of lectures on the basic principles of 4D dilatonic
gravity and some of their consequences for cosmology and
astrophysics. E. Nissimov (Bulgarian Academy of Science) completed the
morning session by presenting his work on strings and branes with
dynamically generated tension. The properties of a new class of string
and p-brane models free of any ad hoc dimensionful parameters were
discussed.  

P. Bozhilov (Shoumen U) started the afternoon session by
presenting his work on probe brane dynamics in nonconstant background
fields. A probe p-brane dynamics in string theory backgrounds of
general type was considered, with an action, which interpolates
between Nambu-Goto and Polyakov type actions. In his talk, B. Ivanov
(Bulgarian Academy of Science) acquainted the audience with his work
on the maximal bounds on the surface redshift of anisotropic stars. It
was shown that for realistic anisotropic star models the surface
redshift cannot exceed values, higher than 2, the bound in the perfect
fluid case, when the tangential pressure satisfies the strong or the
dominant energy condition, respectively. R. Rashkov (Sofia U)
completed the program for the day, by presenting his talk on the
physical states in vacuum string field theory.  

In a spark of
scientific enthusiasm, the Conference continued its work on
Saturday. P. Fiziev (Sofia U) and R. Rashkov (Sofia U) completed their
series of lectures and V. Gueorguiev discussed the subject of
relativistic particle and its D-brane cousin.  The properties of
classical reparametrization-invariant matter systems, mainly the
relativistic particle and its generalization to extended objects
(D-branes) were presented.  

All the organizers were very much satisfied
with the overall results from the Conference. We concluded that there
is a need for such meetings and started preparation for the next
one. One of the main supporting arguments was the very convenient
Sofia University Summer House, which hosted the Conference. In
accordance with the suggestions from the participants, we are
considering building a new Lecture Hall, which will be able to host
bigger crowds. So, we hope to see you next year at the Black Sea.

\vfill\eject
\section*{\centerline {
Quantum field theory on curved spacetime}\\
\centerline{at the Erwin Schr\"odinger Institute}}
\addcontentsline{toc}{subsubsection}{\it
Quantum field theory at ESI, by Robert Wald}
\begin{center}
Robert Wald, University of Chicago
\mbox{\ }
\htmladdnormallink{rmwa@midway.uchicago.edu}
{mailto:rmwa@midway.uchicago.edu}

\end{center}

A program on ``Quantum field theory on curved space times'' was held
at the Erwin Schroedinger Institute in Vienna, Austria from July 1
through August 31, 2002. The main goal of this program was to bring
together researchers with expertise in general relativity and
researchers with expertise in mathematical aspects of quantum field
theory, in order to address some problems of mutual interest in
quantum field theory in curved spacetime.  Approximately 25
researchers in quantum field theory in curved spacetime and related
areas participated in the program. The following is a brief summary of
some of the main topics and results discussed during the program.

A great deal of progress has been made in recent years in
characterizing the ``ultraviolet divergences'' of quantum fields in
curved spacetime and developing renormalization theory for interacting
quantum fields. Seminars by S. Hollands, K. Fredenhagen, R. Verch, and
R. Wald reported on this recent progress. The difficulties resulting
from the lack of a preferred vacuum state and a preferred Hilbert
space representation of the canonical commutation relations for the
free field have been overcome by formulating the theory within the
algebraic approach.  The difficulties associated with the lack of a
global notion of a Fourier transform (so that the usual momentum space
methods for renormalization cannot be used) have been overcome by the use
of the methods of ``microlocal analysis''. Finally, the difficulties
associated with the absence of a notion of ``Poincare invariance'' (or
any other symmetries) in general curved spacetime have been overcome
by imposing the condition that the quantum fields of interest be
constructed locally and covariantly out of the spacetime metric. The
upshot is that perturbative renormalization theory for quantum fields
in curved spacetime is now on as sound a footing as in Minkowski
spacetime. Furthermore, theories that are renormalizable in Minkowski
spacetime will also be renormalizable in curved spacetime, although
additional ``counterterms'' corresponding to couplings of the quantum
field to curvature will arise. 

Although the Hawking effect was derived more than 25 years ago, there
remains a difficulty with the derivation in that it relies on the
properties of quantum fields in a regime where one has no right to
expect quantum field theory in curved spacetime to be a good
approximation. Specifically, consider the modes of the quantum field
that correspond to ``particles'' that are seen by observers near
infinity to emerge from the black hole at late times.  When traced
backward in time, these modes become highly blueshifted and correspond
to ``trans-Planckian'' frequencies and wavelengths at early
times. Thus, the Hawking effect appears to rely on assumptions
concerning the initial state and behavior of degrees of freedom in the
trans-Planckian regime. Similar issues also arise in cosmology when
considering the ``quantum fluctuations'' responsible for the formation
of large scale structure at late times. Seminars by Jacobson and Unruh
explained the nature of the trans-Planckian issues and described some
simple models where the effects of modifying dynamical laws in the
trans-Planckian regime can be analyzed. These models support the view
that the Hawking effect is robust with respect to changes in physical
laws in the trans-Planckian regime.

It is well known that in quantum field theory in flat or curved
spacetime, the expected energy density at a point can be made
arbitrarily negative. However, during the past ten years, some global
restrictions on negative energy have been derived. In particular,
``quantum inequalities'' have been derived, which put a lower bound on
the energy density measured along the worldline of an observer with a
(smooth, compact support) ``sampling function'' $f(\tau)$. Originally,
such bounds were derived by non-rigorous methods in certain special
cases, but recently a rigorous and completely general derivation of
quantum inequalities has been given using the methods of microlocal
analysis. Many issues remain open, however, such as the derivation of
optimal bounds and whether some version may hold of the average null
energy condition (which asserts the non-negativity of the integral
over a complete null geodesic of the stress energy tensor contracted
twice with the tangent to the null geodesic). These issues were
explored in seminars by Ford, Fewster, Roman, Flanagan, and
Pfenning. In research arising directly from discussions occurring
during the program, progress also was made toward deriving quantum
inequalities for quantities other than the stress-energy tensor.

The Bisognano-Wichmann theorem states that in Minkowski spacetime, the
restriction of the vacuum state to a wedge region is a KMS state with
respect to a 1-parameter subgroup of the Poincare group. (This result
can be viewed as a mathematically rigorous version of the ``Unruh
effect''.) The mathematical theory underlying this result is the
modular theory of Tomita and Takesaki.  The computation of modular
transformations was discussed in a seminar by Ynvason, and the
determination of analogous wedge regions and states in deSitter
spacetime was discussed by Guido. In Anti-de Sitter spacetime the
wedges are in one to one correspondence with double cones in the
Minkowski spacetime at spacelike infinity. One thus obtains an
algebraic version of AdS-CFT correspondence which was discovered by
Rehren. A seminar by Rehren discussed this correspondence in more
detail by considering the relation between limits of fields at the
boundary and the partition function for specified boundary values.

In the loop variables/quantum geometry approach to quantum gravity,
one first defines a ``kinematical Hilbert space'' and then tries to
define the action of the Hamiltonian constraint operator on these
``kinematical states''. In this approach, the Hamiltonian constraint
operator is not intrinsically well defined (i.e., ``regularization''
is needed), but the nature of this regularization appears to be very
different from the usual regularization of ``ultraviolet divergences''
occurring in quantum field theory. One of the goals of our program was
to explore the nature of renormalization in the loop variables/quantum
geometry approach to quantum gravity and to understand its
relationship to renormalization in ordinary quantum field
theory. Seminars by Lewandowski, Perez, Ashtekar, Thiemann, Bojowald,
Fairhurst, and Sahlmann described in detail various aspects of the
loop variables/quantum geometry approach. The extended interactions
between the researchers in the loop variables/quantum geometry
approach and researchers in quantum field theory resulting from these
seminars as well as from numerous private discussions were very
fruitful. In particular, simple quantum field theory analogs of some
of the constructions used in the loop variables/quantum geometry
approach were obtained and explored.

Overall, the program appears to have been very successful in promoting
considerable productive interaction between groups of researchers who
generally have had only limited interaction with each other. One may
hope that the ``cross-fertilization'' and new collaborations initiated
by these interactions will bear fruit for many years to come.

\vfill\eject
\section*{\centerline {
School on quantum gravity in Chile}}
\addcontentsline{toc}{subsubsection}{\it
School on quantum gravity in Chile, by Don Marolf}
\begin{center}
Don Marolf, Syracuse University 
\mbox{\ }
\htmladdnormallink{marolf@phy.syr.edu}
{mailto:marolf@phy.syr.edu}

\end{center}

For ten days in January, nearly 70 grad students and postdocs representing
over 13 countries gathered at the Centro Estudios Cientificos (CECS)
in Valdivia, Chile for the PASI school on
quantum gravity.  Supported by the U.S. NSF and DOE through a grant
to Syracuse University, by Syracuse University funds, by the Millennium
Science Initiative (Chile), and by a group of private companies that
supports CECS, the school featured a series of lectures by renowned
experts with ample time for questions and discussion.  My co-organizer
Andres Gomberoff and I were pleased to arrange lectures by
Marc Henneaux, Ted Jacobson, Clifford Johnson, Juan Maldacena, Rob Myers, 
Washington Taylor, Claudio Teitelboim, Bob Wald, and Frank Wilczek. We
were also looking forward to lectures by Abhay
Ashtekar and Rafael Sorkin and, although both had to cancel due
to last minute emergencies, both remained connected to the program.
Sorkin was able to send a set of typed lecture notes that were
widely read and discussed and questions and comments from
several of Ashtekar's students ensured that their point of view
was represented.

Though perhaps we are somewhat biased, 
Andres and I thought that the interactions between the speakers was
excellent -- in particular, between the string theorists and other 
quantum gravity speakers.    Along with the excellent questions from students, 
public discussion among the lecturers helped to draw out the important
but subtle points in each of the talks.
Readers curious about the topics can still find the
schedule posted at 
\htmladdnormallink{http://physics.syr.edu/\~{}pasi/sched.html}
{http://physics.syr.edu/\~{}pasi/sched.html}, but
suffice it to say that the range was broad and the quality was
excellent.  I am very much looking forward to the proceedings that
speakers have promised to write (hint, hint!).  They will no doubt
become a valuable reference for the community.

We were also extremely please with how well the students mixed,
though it is unclear  whether this was due to the intellectual environment, 
the pool parties at the hotel, 
the local Cuban Salsa club, or the sheer joy of the North Americans
experiencing summer in January.  The contacts made with those from other
countries or other perspectives on quantum gravity should serve these
students well throughout their career.
We are very grateful to CECS for providing a dynamic, stimulating, 
and lovely environment for the
school.  Thanks again to everyone who helped to make the event
a success!!

\vfill\eject
\section*{\centerline {
``Apples with Apples no Oranges''} \\
\centerline{Workshop on Formulations of Einstein's Equations}\\
\centerline{for Numerical Relativity}} 
\addcontentsline{toc}{subsubsection}{\it
Apples with apples workshop, by Miguel Alcubierre}
\begin{center}
Miguel Alcubierre, UNAM, Mexico
\mbox{\ }
\htmladdnormallink{malcubi@nuclecu.unam.mx}
{mailto:malcubi@nuclecu.unam.mx}

\end{center}

Late Spring saw a very successful workshop concentrating on
formulations of the Einstein equations for numerical relativity.  The
workshop was held at Institute of Nuclear Sciences of the National
University of Mexico (UNAM) in Mexico city on May 13-24, and was attended by some
25 people from Mexico, the U.S. and Europe (with a particularly large
group from Germany).  The purpose of the workshop was to gather a
group of experts in the recent developments of the different
formulations of the Einstein equations and their applications to
numerical relativity.  There was a focus in 3+1 formulations, though
the conformal approach was also represented.  One of the highlights of
the meeting was our trip to Teotihuacan, where many a pyramid was
climbed and where the god of rain was kind on us by throwing
torrential rains during our bus trip to refresh the place, and
stopping just as we arrived.  But we also talked about physics, of
course.

The workshop was motivated by the growing realization in the numerical
relativity community that different forms of writing the evolution
equations can have tremendous impact in the long term stability and
accuracy of a numerical simulation.  Some formulations have been show
to be either simply ill-posed in a mathematical sense, or else such
that well-posedness can not be proved.  Such is the case, for example,
of the standard ADM formulation.  Formulations that can be shown to be
well-posed, for example strongly hyperbolic ones, should clearly have
an advantage.

During the workshop several key points became clear: hyperbolicity,
though desirable, is not enough, as several groups reported that some
strongly hyperbolic formulations are far superior than others.  Also,
hyperbolicity might not even be necessary for long-term stability, as
some formulations that are not directly hyperbolic (as the BSSN
formulation) have been shown to be remarkably robust.  Moreover,
recent developments, in particular the introduction of
multiple-parameter families of strongly hyperbolic formulations by
Kidder et all (the KST family), make it clear that simply comparing
many formulations against each other in direct numerical experiments
is a difficult task, and theoretical insights are badly needed to help
us chose the more promising formulations from an everyday growing zoo.

The meeting lasted two weeks, and was organized with talks in the
morning and working sessions in the afternoon during the first week,
and informal discussions plus more working sessions the second week.
It is difficult to discuss all contributions to the workshop here,
but I will make an attempt to mention a few highlights:

On Monday, Lee Lindblom talked about the recent KST family of strongly
hyperbolic formulations, and mentioned what I believe to be a very
important development in the field, the realization that
since symmetric hyperbolic formulations allow a norm to be constructed,
one can in fact predict ahead of time the rate of growth of solutions
on a given background.  This has led to the discovery that this growth
is essentially a linear phenomenon, so analysis of linearized
equations around non-trivial backgrounds should be sufficient.  Carles
Bona then presented a framework under which many different formulations,
from the ADM and BSSN formulations to the Bona-Masso and KST
formulations can be studied under a unified approach.

On Tuesday, Hisaaki Shinkai showed the results of many tests he has
carried out with many formulations, and presented the hypothesis that
a Fourier mode analysis of linearized equations around specific
backgrounds should be a good indicator of which formulation will
behave better than others.  Thomas Baumgarte presented a beautiful
analysis of electrodynamics where he showed how one can construct an
analogous of the BSSN formulation in relativity and see very clearly
what the benefits of such a form of the evolution equations are.  Erik
Schnetter talked about ways in which different gauge conditions can
and should be enforced during a numerical simulation.  Finally,
Deirdre Shoemaker talked about recent advances that Pablo Laguna's
group at PSU has had in developing formulations based on BSSN, but
where care is taken to eliminate some quadratic source terms.  Their
approach is to look systematically not only at the principal part, but
also a lower order terms in the search for better stability.

Wednesday saw the talks representing the conformal approach, given by
Christiane Lechner and Sascha Husa.  Their approach, based on
Friedrich's conformal equations is capable of evolving all the way to
null infinity (and beyond), seems to have all the benefits of the
Cauchy-characteristic matching and fewer of its problems.  Still, at
this time it also seems to have problems with long term stability.
Wednesday also saw our trip to Teotihuacan.

On Thursday Jeff Winicour and Bela Szilagyi reminded us all of the
importance of choosing adequate boundary conditions that are designed
to satisfy the constraint equations.  Their efforts on developing
consistent boundary conditions are a model of what all other groups
should be worrying about when writing Cauchy codes.  The interaction
of boundaries and interior should clearly be taken into account when
thinking about the long-term stability of the different codes.  Ryoji
Takahashi also talked about gauge stability of different 3+1
formulations.

Friday saw a talk by Miguel Alcubierre presenting the stability
analysis of BSSN in linearized gravity.  Denis Pollney talked about
the recent experience of the Potsdam group with BSSN in black hole
simulations, and Ian Hawke gave a general talk on how to implement in
a simple and systematic way a general method of lines algorithm
within the Cactus framework.

Working sessions concentrated in getting the different codes working
together under a similar framework.  Whenever possible, different
codes where ported into the Cactus framework to facilitate sharing
initial data and analysis tools.  The meeting was very successful, and
people have continued to work together several months afterward
trying to compare results of the different approaches in specific
situations.

As a final personal comment I would like to add that the workshop had
also an impact in the Institute of Nuclear Sciences in Mexico, where I
work.  Initially, a computer terminal room was booked for the working
sessions, but it quickly became apparent that laptops and wireless
networks could be used to transform our main lecture theater into a
much better working environment.  The sight of 20 plus people having
claimed the main auditorium for two weeks solid, typing away at their
laptop keyboards while simultaneously sharing data through the Internet
was enough to impress everyone, and in my view showed what should
probably become a standard in future numerical relativity workshops.

\vfill
\eject
\section*{\centerline {
Radiation reaction focus session} \\
\centerline{and the 5th Capra meeting}}

\addcontentsline{toc}{subsubsection}{\it 
Radiation reaction focus session and 5th Capra meeting,
by \'Eanna Flanagan}
\begin{center}
    \'Eanna Flanagan, Cornell \\
\htmladdnormallink{eef3@cornell.edu}
{mailto:eef3@cornell.edu}
\end{center}
\parindent=0pt
\parskip=5pt

One of the activities supported by the newly created Center for
Gravitational Wave Physics at Penn State are {\it focus sessions}.
These are intended to bring together a small group of experts in a single,
narrowly defined technical topic in order to make progress (see the
article by Sam Finn in Issue 19 of Matters of Gravity).  
One of the first such focus sessions was held at Penn State from May
24 to 30, dedicated to radiation reaction in general relativity,
organized by Warren Anderson, Patrick Brady, Sam Finn and myself.
It was followed by the fifth Capra Ranch meeting on radiation
reaction, from May 31 to June 2.

\medskip
\centerline
{\it Focus session on radiation reaction}

One of the primary sources for the space-based gravitational wave
detector LISA will be the inspiral of compact objects (neutron stars
and solar mass black holes) into supermassive black holes.  
The last year of inspiral of these systems will typically consist of
$\sim 10^5$ or so orbits deep in the relativistic regime near the
black hole horizon (Finn and Thorne 2000).  In order to compute the
gravitational wave signal emitted by these systems, it is hoped to
compute the deviations from geodesic motion 
to linear order in the mass ratio, using black hole perturbation theory.  
While conservation laws can be used to evolve certain classes of
orbits (see, eg, Hughes 2000), the treatment of generic orbits will
require the computation of the so-called self force that acts on the
compact object.

Over the last few years, several research groups have been developing
the mathematical and computational tools necessary for computation of
the self force; see the accounts of the previous Capra Ranch meetings
in Issues 14, 16 and 18 of Matters of Gravity.  The foundations for
this effort were laid by a pair of papers in 1997 (Mino, Sasaki and
Tanaka 1997; Quinn and Wald 1997), which derived a general formal 
expression for the self force in an arbitrary vacuum spacetime.  The
challenge has been to translate this expression into a practical
computational scheme for orbits in the Kerr spacetime.  
Such a computational scheme may also be useful for 
compact objects spiralling into middleweight black holes ($\sim 10^3
M_\odot$) which may be detected by ground based detectors (see the
article by Ben Bromley in Issue 14 of Matters of Gravity).  
While several different computational schemes are being explored, the
most developed and most promising contender at the moment seems to be
``mode sum regularization'', in which one (i) computes using black
hole perturbation theory the contribution $f^\alpha_{lm\omega}$ to the
self force on a point particle from the mode $lm\omega$; (ii)
subtracts from this  an analytically computed counterterm involving
some regularization parameters; and (iii) sums over all the modes to
determine the total self force (see, eg, Barack et. al. 2002 and
references therein).  This scheme has by now been 
implemented to 
compute gravitational self-forces in the Schwarzschild spacetime
(Barack and Lousto 2002).  The challenge now is to extend the analysis
to the Kerr spacetime.

It was to address this challenge that Sam Finn instigated the focus
session.  We were fortunate to be able to attract most of the key
researchers in the field to attend; eighteen people in all were given
office space and internet connections for a week.  
The format and organization of the focus session was significantly
different from traditional conferences based around presentations,
and was aimed instead at promoting interactions and collaborations
among the participants.  While one or two black board 
talks/discussions were scheduled most days, a significant amount of
time was left open.

Some of the highlights of the week were as follows.  On the first day
Yasushi Mino gave a detailed overview of the status of self-force
computations, and Scott Hughes reviewed LISA Science objectives.
One of the key technical issues that dominated the discussions of the
week was the incomplete status of the theory of linear perturbations
of the Kerr spacetime.  Specifically, one needs to be able to
reconstruct the metric perturbation from the Weyl scalars $\psi_0$ and
$\psi_4$ for general non-vacuum linear perturbations.  Bernard Whiting
discussed a proposed method of completing the formalism of Chrzanowski
(1975) as corrected by Wald (1978), and Amos Ori discussed another
possible method.  Steve 
Detweiler talked about how one might deal with the fact that 
the Teukolsky-Sasaki-Nakamura perturbation formalism does not include
the the ``$l=0,1$'' modes which are needed for self-force
computations.  Another theme was the freedom of choice in the
analytically computed counterterms that one uses to renormalize the
self force due to a particular mode.  
Leor Barack discussed the
details of one choice of counterterms, and Steve Detweiler
discussed another choice that increased the speed of convergence in the
sum over modes (Detweiler, Messaritaki and Whiting 2002) based on a
definition of a regularized self-field that satisfies the homogeneous
wave equation (Detweiler and Whiting 2002).

Eric Poisson discussed the necessity of going beyond the computation
of the self-force to calculate the gravitational wave signal, and the problem
that linear perturbation theory is strictly speaking insufficient to
compute the gravitational wave signal from the inspiralling orbit that
includes the backreaction.  For the most general contexts, it will be
necessary to use second order perturbation theory (Campanelli and
Lousto 1999), a daunting prospect!  However, in the adiabatic inspiral
regime relevant to most of the observations, it seems likely that the linear
perturbation formalism will be highly accurate, and there was some
discussion of how one might justify this using a two-timescale
expansion of the Einstein equations rather than a straightforward
perturbation expansion.

One particularly useful talk was Bob Wald's review of the
construction of Green's functions via Hadamard expansions, in which he
debunked several myths that have appeared in the relativity
literature.  In particular, the radius of convergence of the Hadamard
series can be zero for smooth Lorentzian metrics, and outside of a
normal neighborhood one cannot in general compute a Green's function
$G(x,x^\prime)$ by summing over the geodesics that join $x$ and $x'$.  
Eric Poisson discussed how the Mino-Sasaki-Tanaka-Quinn-Wald
self-force expression, when specialized to weak fields and slow
motions, reproduces the standard post-1-Newtonian results wherein the
self force has a conservative part but no dissipative part (Pfenning
and Poisson 2002).  
A separate approach to computing self-forces would be to use a numerical,
time-domain Teukolsky code rather than splitting the field into
modes.  Carlos Lousto reviewed the status of this field of research.
Warren Anderson discussed computations with Adrian Ottewill and myself
of a local expansion of the tail piece of the metric perturbation,
which could be used as a foundation for a different type of
regularization method.

There was an enthusiastic consensus by the end of
the week that the format had worked very well, and that the intensive
discussions that grew out of the blackboard talks were very useful. 
These discussions spilled over into the coffee breaks, the offices of
the participants, and to the picnic hosted by Sam Finn one evening. 
The meeting was particularly useful because the people
involved had already formed a small, closely knit group via the
previous Capra Ranch meetings, and most of the group were focusing closely
on a specific narrowly defined research area.  I expect that many of the
discussions at the sixth Capra meeting next summer will involve
research projects and collaborations that were germinated at the focus
session.

\medskip
\centerline
{\it Fifth Capra Ranch Meeting}

The primary focus of the meeting was radiation
reaction of point particles, as in previous years, but there was also 
broader, forward-looking focus on radiation reaction effects in
general and in particular in numerical relativity and astrophysics.
We followed the informal format used in previous meetings where equal
time was allotted to talks and discussions.

The meeting opened with a presentation by Amos Ori on a suggested
method for computing the metric perturbation in the ingoing radiation
gauge from the Weyl scalars $\psi_0$ and $\psi_4$ in the Kerr
spacetime, based on working in the frequency domain.  He described
how this method could be used in principle to compute self forces
(see Ori 2002 for details).  Yasushi Mino described work in progress
on a different approach towards computing the self force in Kerr,
using the Regge-Wheeler gauge and an expansion in powers of the black
hole's spin parameter $a$.  

Leor Barack and Carlos Lousto discussed different aspects of their recent
and ongoing work on
computing gravitational self forces in the Schwarzschild spacetime
(Barack and Lousto 2002),
based on the method of Barack {\it et.\ al.\ } (2002).
They described the analytic computation of all the regularization
parameters needed for a generic orbit, completed computations for
radial trajectories that agreed with the $\zeta$-function regularization
scheme of Lousto (2000), and ongoing work on circular orbits.  Leor also
described analytic computations of the large $l$ behavior of the
un-renormalized self force; this information was used to improve the
speed of convergence of the sum over modes.

Lior Burko described computations of scalar field radiation reaction
for particles in circular orbits about Schwarzschild black holes
(Burko 2002), which incorporated corrections to the
phasing beyond the leading order in an expansion in the mass ratio
(corresponding to an accumulated phase correction of order unity during
an inspiral).   He discussed the fact that such higher order
corrections may eventually be necessary for precision astronomy with
LISA.

Steve Detweiler described the Green's function decomposition of
Detweiler and Whiting (2002).  The advantages of this decomposition
over earlier decompositions is that the corresponding regularized metric
perturbation is a smooth solution of the homogeneous wave equation,
and the particle's motion is a geodesic of the regularized perturbed
metric.  He also described how the use of a particular normal
coordinate system defined by Thorne and Hartle (1985) and extended by
Zhang (1986) greatly simplifies the local computation of the singular
piece of the metric perturbation that is subtracted, and related the
construction to the derivation of the self force based on matched
asymptotic expansions given in Detweiler (2001).  He then described
scalar self-force computations for particles in 
Schwarzschild (Detweiler, Messaritaki and Whiting 2002)
and also new gravitational self-force computations, and showed
how the use of the new decomposition speeded 
up the convergence of the sum over modes.

Saturday started with Eric Poisson describing ongoing work with Claude
Barrab\`es 
in which he defined a coordinate system called retarded normal
coordinates, which are closely related to Fermi normal coordinates
except that the construction is based null geodesics rather than
spacelike geodesics.  He showed how the use of these coordinates
rather than other types of normal coordinates greatly simplifies the
computations in the classic papers of Dirac (1938) and DeWitt and
Brehme (1960), and he anticipated that they would also simplify
the computations of Mino, Sasaki and Tanaka (1997) and Quinn and Wald
(1997).

Next, Manuela Campanelli described a formalism for computing second order
perturbations of Kerr black hole in the time domain using the Weyl
scalar $\psi_4$ (Campanelli and Lousto 1999).  She described how to
obtain second order quantities that are invariant under both
coordinate transformations and tetrad 
rotations, and showed impressive numerical results comparing the
second order gravitational waveform from binary black holes in the
close limit with the waveform from full numerical relativity and with
the first order waveform.  She explained that the main roadblock to
applying the 
code to compute self-forces is the necessity to regularize the
formally infinite source terms representing the point particle.  
On a similar note, Karl Martel described numerical work in which he
explored replacing a delta function source with a Gaussian profile, in
the context of computing the scalar field sourced by a point particle in 
Schwarzschild.  He compared the waveform obtained from the smeared
source to that from the exact source for a point particle, for
computations in both Schwarzschild coordinates and  
Painlev\'e-Gullstrand coordinates.  He showed that good agreement
between the waveforms was obtained when the width of the Gaussian
profile was suitably chosen.

Bernard Whiting then gave a talk in which he discussed a number of
issues, including 
the status of the problem of the $l=0,1$ modes in Kerr, the reason why
increased smoothness of the regularized metric perturbation gives rise
to improved convergence properties of the sum over modes, and
analogous phenomena involving convergence and smoothness that arise
in LIGO data analysis.  He also discussed the prospects for
generalizing to the Kerr spacetime the formalism of Lousto and Whiting
(2002) for reconstructing metric perturbations from Weyl curvature
perturbations in Schwarzschild, in which he emphasized the key role
of the algebraically special solutions.

Ian Jones described an ongoing project at Southampton University
aimed at including local radiation reaction forces in nonlinear
Newtonian and post-Newtonian hydrodynamic codes.  He discussed
numerical difficulties involved involved in evaluating the local force
expressions caused by the large number of time derivatives required.
He reviewed a particular formulation of post-1-Newtonian hydrodynamics
due to Blanchet, Damour and Schaeffer that eliminates the time
derivatives, and which is well adapted to mass quadrupole radiation
reaction, and indicated that they planned to use an extension of this
formalism being developed by Faye and Schaeffer to incorporate current
quadrupole radiation reaction.

Mark Miller described fully general relativistic, 3+1 dimensional
simulations of binary neutron star inspirals, using the code described
in Font et.\ al.\ (2001), and compared the orbital decay rate obtained
to the decay rates obtained from post-Newtonian computations.
The full GR code evolved the binary for 10 orbital periods.  He explained that
oscillations observed in the stellar separation suggested that the initial
data used actually corresponded to a slightly eccentric binary.  He
also showed how to use Richardson extrapolation with several 
runs with different outer boundaries and grid sizes to estimate the 
computational error.  Currently the error in the decay rate is
comparable to the decay rate itself, but the errors will improve with
time.  He also described computations of binding energy curves
for binaries obtained using the so-called conformally-flat, quasi-equilibrium
approximation, and the orbital decay rates obtained by combining those
binding energy curves with the quadrupole energy loss rate.  To
compute the binding energy curves, he advocated a new prescription in
which one subtracts from the total ADM mass of the spacetime the sum of the
ADM masses of isolated, {\it rotating} neutron stars with
appropriately chosen angular momenta (rather than
isolated non-rotating neutron stars as had been done in the past).

The Capra meeting, like the focus session, was supported by funds
from the Center for Gravitational Wave Physics.  Thanks are due to Sam
Finn for his organizational skills.  All of the
presentations can be found online at 

\htmladdnormallink{http://cgwp.gravity.psu.edu/events/Capra5/capra5.html}
{http://cgwp.gravity.psu.edu/events/Capra5/capra5.html}.

\bigskip

{\bf References:}

Barack, L., and Lousto, C., 2002,
\htmladdnormallink{gr-qc/0205043}{http://arXiv.org/abs/gr-qc/0205043}.
\hfil\break
%
%
Barack, L., Mino, Y., Nakano, H., Ori, A., and Sasaki, M., 2002,
Phys. Rev. Lett. {\bf 88}, 091101 
(also \htmladdnormallink{gr-qc/0111001}{http://arXiv.org/abs/gr-qc/0111001}).
\hfil\break
%
%
%
Burko, L., 2002, 
\htmladdnormallink{gr-qc/0208034}{http://arXiv.org/abs/gr-qc/0208034}.
\hfil\break
%
%
Campanelli, M., and Lousto, C., 1999, Phys. Rev. D {\bf 59}, 124022
(also
\htmladdnormallink{gr-qc/9811019}{http://arXiv.org/abs/gr-qc/9811019}).
\hfil\break
%
%
Chrzanowski, P.L., 1975, Phys. Rev. D {\bf 11}, 2042.
\hfil\break
Detweiler, S., Messaritaki, E., and Whiting, B., 2002, 
\htmladdnormallink{gr-qc/0205079}{http://arXiv.org/abs/gr-qc/0205079}.
\hfil\break
%
%
Detweiler, S., and Whiting, B., 2002,
\htmladdnormallink{gr-qc/0202086}{http://arXiv.org/abs/gr-qc/0202086}.
\hfil\break
%
%
DeWitt, B.S., and Brehme, R.W., 1960, Ann. Phys. {\bf 9}, 220.
\hfil\break
Dirac, P.A.M., 1938, Proc. R. Soc. (London) {\bf A 167}, 148.
\hfil\break
Finn, L.S., and Thorne, K.S., 2000, Phys. Rev. D {\bf 62}, 124021
(also \htmladdnormallink{gr-qc/0007074}{http://arXiv.org/abs/gr-qc/0007074}).
\hfil\break
%
%
%
Font, J.A. {\it et. al.}, 2002, Phys. Rev. D {\bf 65}, 084024
(also \htmladdnormallink{gr-qc/0110047}{http://arXiv.org/abs/gr-qc/0110047}).
\hfil\break
Hughes, S., 2000, Phys. Rev. D {\bf 61}, 084004 (also 
\htmladdnormallink{gr-qc/9910091}{http://arXiv.org/abs/gr-qc/9910091}); 
Errata Phys Rev D {\bf 63}, 049902 (2001) and {\bf 65}, 069902 (2002).
\hfil\break
%
%
Lousto, C.O., 2000, Phys. Rev. Lett. {\bf 84}, 5251 
(also
\htmladdnormallink{gr-qc/9912017}{http://arXiv.org/abs/gr-qc/9912017}).
\hfil\break
%
%
Lousto, C.O., and Whiting, B.F., 2002, Phys. Rev. D {\bf 66}, 024026 
(also
\htmladdnormallink{gr-qc/0203061}{http://arXiv.org/abs/gr-qc/0203061}).
\hfil\break
%
%
Mino, S., Sasaki, M., Tanaka, T., 1997, Phys. Rev. D {\bf 55}, 3547
(also \htmladdnormallink{gr-qc/9606018}{http://arXiv.org/abs/gr-qc/9606018}).
\hfil\break
%
%
Ori, A., 2002, 
\htmladdnormallink{gr-qc/0207045}{http://arXiv.org/abs/gr-qc/0207045}.
\hfil\break
%
%
Pfenning, M.J., and Poisson, E., 2002, Phys. Rev. D {\bf 65}, 084001 
(also
\htmladdnormallink{gr-qc/0012057}{http://arXiv.org/abs/gr-qc/0012057}).
\hfil\break
%
%
Quinn, T.C., and Wald, R., 1997, Phys. Rev. D {\bf 56}, 3381 
(also \htmladdnormallink{gr-qc/9610053}{http://arXiv.org/abs/gr-qc/9610053}).
\hfil\break
%
%
%
%
Thorne, K.S., and Hartle, J., 1985, Phys. Rev. D {\bf 31}, 1815.
\hfil\break
%
Wald, R.M., 1978, Phys. Rev. Lett. {\bf 41}, 203.\hfil\break
%
%
Zhang, X.-H., 1986, Phys. Rev. D {\bf 34}, 991.
\hfil\break

\vfill
\pagebreak

\section*{\centerline {
Numerical Relativity Workshop at IMA}}
\addcontentsline{toc}{subsubsection}{\it 
Numerical Relativity Workshop at IMA, by Manuel Tiglio}
\begin{center}
    Manuel Tiglio, Louisiana State University
\htmladdnormallink{tiglio@lsu.edu}
{mailto:tiglio@lsu.edu}
\end{center}

The {\it Numerical Relativity (NR) Workshop} organized by the
Institute for Mathematics and its Applications (IMA) at the University
of Minnesota and jointly sponsored with the Center for Gravitational
Wave Physics (CGWP) at Penn State, took place in IMA's facilities from
June 24th to June 29th (the last week of the 2002 Soccer World Cup, a
detail not ignored by many of the attendees).

As announced, the workshop effectively ``brought together numerical
relativists and mathematicians working in fields such as numerical
analysis, scientific computation, partial differential equations and
geometry, for an intense but informal period aimed at maximal
communication and interaction between diverse
researchers''. Considerable effort was put in bringing these
communities closer. Relativists tried to describe the status and
problems of NR and differences between General Relativity (GR) and
Maxwell theory or fluid dynamics. On the other hand, numerical
analysts made their best to bring their expertises closer to our
field. At the end of the workshop the latter were much more familiar
with the problems of NR, while relativists got to learn and revisit a
plethora of mathematical techniques that may help solve some of the
problems.

On the first day mathematicians were introduced to 
GR by talks given by Doug Arnold (director of IMA), Alan Rendall
(AEI), and Robert Bartnik (University of Canberra). Arnold gave a self
consistent tutorial 
to Einstein's equations from a mathematician's point of view. Followed
Rendall, discussing  
the $3+1$ decomposition and local existence and uniqueness to the
Cauchy problem. Bartnik, in turn, 
explained slicing conditions, the constraint equations and 
gravitational radiation.

On Tuesday morning, Ralf Hiptmair (Universit\"at Bonn) reviewed
discretizations of Maxwell's equations, putting emphasis on coordinate-
free methods. He included the finite volume approach, generalized finite
differences, and finite elements. Later, Eitan Tadmor (University of Maryland)
presented discretizations for nonlinear hyperbolic systems that preserve local
and global invariants. Since these discretizations are motivated
mainly by computational fluid dynamics, there were several discussions
on how to make use of these techniques in Einstein's vacuum
evolutions. On Thursday morning Tadmor continued the discussion 
with some remarks about hyperbolic formulations and giving some ideas
 to diminish the accumulation errors by trading some hyperbolic
equations for elliptic ones.

Back on Tuesday, during the afternoon Oscar Reula (University of
C\'ordoba) talked about the role of hyperbolicity in formulations of
Einstein's equations. He discussed the weak-ill posedness of the
standard ADM formulation and the ``hidden'' hyperbolicity of BSSN-like
systems. He also mentioned the linear degeneracy of usual formulations
of Einstein's equations.  Markus Keel (IMA) on Thursday morning gave a
talk further elaborating this point and others. Namely, he explained
that usual formulations for Einstein's vacuum equations are not
genuinely nonlinear and therefore one should not expect shocks. He
also discussed the ``stability problem'' in NR (later in the workshop
to be covered in more detail by Lindblom and Scheel), and a stability
result by Kreiss, Ortiz and Reula for hyperbolic systems. Going back
once again to Tuesday, after Reula's talk I gave one overviewing NR as
an initial-boundary value problem, summarizing what is known about
constraint preservation, well posedness, and numerical stability for
finite difference schemes in the presence of boundaries.

On Wednesday morning, Jeff Winicour (University of Pittsburgh) gave a
one hour and a half introduction to black holes (BHs), covering topics
such as conformally compactified spacetimes, the notion of an event
horizon and its intrinsic geometry, the no-hair hypothesis, and a
description of gravitational collapse. Later in the morning, Matt
Choptuik (University of British Columbia) talked about fundamental
issues in NR. Among other things, he discussed BH excision, coordinate
conditions, optimization, differences with other kind of numerical
computations, adaptative mesh refinement, gravitational collapse, and
boundary conditions. He pointed out something that was going to be
mentioned by mathematicians several times during the workshop as
well. Namely, that while the majority of NR implementations are
obtained with free (unconstrained) evolution, there are schemes for
solving elliptic equations (e.g. multigrid methods) where the number
of operations scales linearly with number of gridpoints. As
highlights, he discussed the issue of choosing good model problems,
Brandt's golden rule that {\it the amount of numerical computation
should be proportional to the amount of real physical changes in the
computed system}, and that the situations for which we most need NR
are those for which there is little a priori knowledge.

Pablo Laguna (PSU) gave the last talk on Wednesday afternoon, giving an
overview of the state of the art in NR. He reviewed $3$D
evolutions of single BHs,  BH-BH, BH-neutron stars (NSs), and NS-NS
binaries, and $2$D simulations of gravitational collapse. He explained what
formulations of Einstein's equations are used in these simulations,
and the available results.

Later in the afternoon there was a pizza dinner followed by
computational 
demonstrations. Winicour presented numerical evolutions of a well posed
initial-boundary formulation of vacuum GR using the harmonic gauge,
and some comparison tests
from the {\em Workshop on Formulations of Einstein equations for NR}
\footnote{See the article by Alcubierre in this issue of MOG} that 
showed improved stability in this formulation. 
Jincha Xu (Penn State)  discussed problem-independent
adaptivity and multigrid methods. Michael Holst (University of California at
San Diego) presented the (finite element) Manifold Code (MC) developed
at UCSD, designed
to solve nonlinear elliptic systems. Bartnik showed
quasi-spherical numerical evolutions of BHs via a characteristic formulation. Dennis Pollney
(AEI) presented $3$D 
simulations using new coordinate conditions that they applied to
 binary BH evolutions, with improved stability properties.

Thursday morning was devoted to the initial data problem, with two 
talks, given respectively by Greg Cook (Wake
Forest University) and
Holst. Cook reviewed the 3+1 decomposition, the conformal and physical 
transverse-traceless decompositions, and the conformal 
thin sandwich approach. Then, he described BHs and NSs initial data. 
He ended pointing out current 
trends, such as initial data with more astrophysical content and/or
data for objects
in quasi equilibrim. Holst's talk was a natural continuation of
Cook's, this time addressing mathematical issues. For example, he 
overviewed existence results for the Hamiltonian constraint, as well
as more recent techniques for the coupled Hamiltonian-Momentum
 equations. He gave a detailed discussion of well posedness, 
a priori and a posteriori error estimates, and then moved to numerical solutions, 
mainly using finite element methods. He also presented some 
specific results for BH initial data. 

In the afternoon, Deirdre Shoemaker (PSU) talked about BH excision and
apparent horizon tracking. She discussed different numerical methods
for this tracking, as well as the different discretizations that are
currently used for finite differencing at the inner boundary in the
case of excision. As an application she presented the status of
simulations of moving $3$D Schwarzschild BH. Followed Mark Scheel's
(Caltech) talk about pseudospectral evolution of BHs.  He explained
pseudo spectral collocation methods and then discussed the constraint
violating problem in NR (namely, fast growing, non-physical modes
excited by numerical errors).  He showed several evolutions of a $3$D
Schwarzschild BH, emphasizing the dependence of the stability on the
choice of the formulation of Einstein's equations. He also discussed a
priori analytical energy estimates, a point that was later elaborated
in more detail by Lee Lindblom (Caltech) on Friday. In his talk
Lindblom explained their method to obtain sharp estimates, which can
be used to a priori choose a formulation that optimizes the stability
of a given background, and presented numerical results that confirmed
their analytical predictions.

On Friday morning, Luis Lehner (University of British Columbia) gave
the state of the art of outer boundary conditions in NR. He discussed
standard and new ways of handling boundaries in Cauchy evolution, and
the characteristic and conformal approaches. He also stressed that one
can actually make use of Gustaffson-Kreiss-Sundstrum-Osher's theory in
NR to construct stable boundary treatments. Richard Falk (Rutgers
University) explained finite element methods for hyperbolic equations,
how to parallel continuum estimates in the design of numerical
methods, and different mesh constructions. Sascha Husa (AEI)
summarized the conformal approach to NR, discussing previous work
(e.g. constructing the whole spacetime describing weak gravitational
waves) and some needs (for example, live gauges) and difficulties . He
emphasized that many of these difficulties also appear in other, more
standard, approaches to NR.

Sam Finn and Lee Lindblom closed Friday's presentations with two
talks. During the week, computational scientists had asked how
gravitational waves are related to solutions of Einstein's
equations. Finn addressed these questions with a talk that explained
the basics of laser interferometry, and generation and observation of
gravitational waves. Lindblom talked about simulation of r-modes
through pseudo-Newtonian models.

During the afternoon, there was a panel discussion on numerical
methods. Topics such as adaptative mesh refinement
and the possibility of general purpose tools were discussed. The
possibility of explicitly solving
  the constraints during evolution was also brought up, especially given
that  several of the computational scientists present were
experts in solving elliptic equations with advanced, efficient
methods. The current lack of
understanding in NR of boundary conditions
for hyperbolic-elliptic systems was in turn presented by numerical
relativists as one of the main obstacles in implementing such methods.

The workshop ended on Saturday noon, after a discussion lead by
Richard Price. He presented the ``many aspects of the elephant'', and
an original {\it yin-yang} summary of different opinions about to attempt
obtaining gravitational wave templates in the short term 
(we are doing fine/we need a
new idea, specific focus/all areas, careful analysis/computer
experiments, more intuition/more math, big codes/model problems, etc)
that was used to start a general discussion of what would be the best
way to proceed.  

People agreed on continuing the workshop in the future with other
meetings. Doug Arnold suggested joining together smaller 
groups of people willing to collaborate or to 
discuss more specific aspects. 
Sam Finn, as director of the CGWP, offered the Center as another
natural place for these future meetings.
As one can imagine from the number of talks given and variety of topics
discussed, the workshop was very intense. People went back home 
with new ideas, renewed hopes, and just in time
to see Brazil win the final game of the World Cup. 

The conference website is
\htmladdnormallink{http://www.ima.unm.edu/nr}{http://www.ima.unm.edu/nr}

\end{document}